\documentclass[a4paper,12pt,table,noblocks]{biophys-new}

\usepackage[numbers]{natbib}

\usepackage[utf8]{inputenc}
\usepackage{graphicx}
\usepackage{pifont}
\usepackage{pbox}
\usepackage{pdflscape}
\usepackage{array}
\usepackage{xcolor}
\usepackage{transparent}
\usepackage{wrapfig}
\usepackage{lipsum}

\papertype{Article}

\usepackage{caption}
\usepackage{subcaption}

\usepackage{amsmath,amssymb,amsfonts}
\newcommand{\ipr}{IP\textsubscript{3}R }
\newcommand{\iprns}{IP\textsubscript{3}R}
\newcommand{\iprs}{IP\textsubscript{3}Rs }
\newcommand{\iprsns}{IP\textsubscript{3}Rs}
\newcommand{\ip}{IP\textsubscript{3} }
\newcommand{\ipns}{IP\textsubscript{3}}
\newcommand{\ca}{Ca\textsuperscript{2+} }
\newcommand{\cans}{Ca\textsuperscript{2+}}
\newcommand{\caconc}{[Ca\textsuperscript{2+}] }
\newcommand{\caconcns}{[Ca\textsuperscript{2+}]}

\definecolor{dgreen}{rgb}{0,0.6,0}
\newcommand{\mynotes}[1]{\textcolor{black}{#1}}

\renewcommand{\d}[1]{\ensuremath{\operatorname{d}\!{#1}}}
\newcommand{\deriv}[2]{\frac{\d{#1}}{\d{#2}}}

\usepackage[symbol]{footmisc}

\title{\ca release via \ip receptors shapes the cytosolic \ca transient for hypertrophic signalling in ventricular cardiomyocytes}
\runningtitle{On the role of \ipr in cardiac hypertrophic signalling} %% For page header
\runningauthor{Hunt, Til\=unait\.e et al.} %% For page header

\author[1]{Hilary Hunt}
\author[1]{Agn\.e Til\=unait\.e} 
\author[1]{Greg Bass}
\author[2]{Christian Soeller}
\author[3]{H. Llewelyn Roderick}
\author[4]{Vijay Rajagopal\raisebox{2pt}{\small{*}}\footnote[2]{These authors contributed equally to the supervision of this work.}}
\addtocounter{footnote}{-1}
\author[1,5]{Edmund J. Crampin\raisebox{2pt}{\small{*}}{\textsuperscript{$\dagger$}}}

\affil[1]{Systems Biology Laboratory, School of Mathematics and Statistics and Melbourne School of Engineering, University of Melbourne, Australia}
\affil[2]{Living Systems Institute, University of Exeter, UK}
\affil[3]{Laboratory of Experimental Cardiology, Department of Cardiovascular Sciences, KU Leuven, Belgium}
\affil[4]{Cell Structure and Mechanobiology Group, Department of Biomedical Engineering, Melbourne School of Engineering, University of Melbourne, Australia}
\affil[5]{ARC Centre of Excellence in Convergent Bio-Nano Science and Technology, School of Chemical and Biomedical Engineering, University of Melbourne, Australia}
\corrauthor[*]{vijay.rajagopal@unimelb.edu.au; edmund.crampin@unimelb.edu.au}

\begin{document}

\begin{frontmatter}

\begin{abstract}
Calcium (Ca$^{2+}$) plays a central role in mediating both contractile function and hypertrophic signalling in ventricular cardiomyocytes. L-type \ca channels trigger release of Ca$^{2+}$ from ryanodine receptors (RyRs) for cellular contraction, \mynotes{while signalling downstream of Gq coupled receptors stimulates \ca release via inositol 1,4,5-trisphosphate receptors (\iprsns), %modulating the excitation-contraction coupling (ECC) signal and 
engaging hypertrophic signalling pathways}. Modulation of the amplitude, duration, and duty cycle of the cytosolic Ca$^{2+}$ contraction signal, and spatial localisation, have all been proposed to encode this hypertrophic signal. 
%Given current knowledge of \iprsns, we we use mathematical modelling to investigate a possible mechanism , cardiac ECC machinery, and the sensitivity to Ca$^{2+}$ of downstream proteins in the hypertrophic signalling pathway. 
Given current knowledge of \iprsns, we develop a model describing the effect of functional interaction (cross-talk) between RyR and \ipr channels on the Ca$^{2+}$ transient, and examine the sensitivity of the Ca$^{2+}$ transient shape to properties of \ipr activation. A key result of our study is that \ipr activation increases Ca$^{2+}$ transient duration for a broad range of \ipr properties, but the effect of \ipr activation on Ca$^{2+}$ transient amplitude is dependent on \ip concentration. Furthermore we demonstrate that \ipns-mediated \ca release in the cytosol increases the duty cycle \mynotes{of the Ca$^{2+}$ transient, the fraction of the cycle for which \caconc is elevated,} across a broad range of parameter values and \ip concentrations. \mynotes{When coupled to a model of downstream transcription factor (NFAT) activation, we demonstrate that there is a high correspondence between the Ca$^{2+}$ transient duty cycle and the proportion of activated NFAT in the nucleus.} These findings suggest increased cytosolic \ca duty cycle as a plausible mechanism for \ipns-dependent hypertrophic signalling via Ca$^{2+}$-sensitive transcription factors such as NFAT in ventricular cardiomyocytes.
\end{abstract}

\begin{sigstatement}
Many studies have identified a role for IP$_{3}$R-mediated Ca$^{2+}$ signalling in cardiac hypertrophy, however the mechanism by which this signal is communicated within the cardiomyocyte remains unclear. We present a mathematical model of functional interactions between RyR and IP$_{3}$R channels. We show that IP$_{3}$-mediated Ca$^{2+}$ release is capable of providing a modest increase to the duty cycle of the calcium signal, which has been shown experimentally to lead to NFAT activation, and hence hypertrophic signalling. Through a parameter sensitivity analysis we demonstrate that the duty cycle is increased with IP$_{3}$ over a broad parameter regime, indicating that this mechanism is robust\mynotes{, and we show that an increase in Ca$^{2+}$ duty cycle increases nuclear NFAT activation.} These findings suggest a plausible mechanism for IP3R-dependent hypertrophic signalling in cardiomyocytes.
\end{sigstatement}

\end{frontmatter}

\section*{Introduction}
Calcium is a universal second messenger that plays a role in controlling many cellular processes across a wide variety of cell types; ranging from fertilisation, cell contraction, and cell growth, to cell death \cite{berridge_calcium_2003,clapham_calcium_2007}. Precisely how Ca$^{2+}$ fulfills each of these roles while also ensuring signal specificity remains unclear in many cases. Ca$^{2+}$ can be used to transmit signals in a variety of ways. Signal localisation, and amplitude and frequency modulation have been widely explored \cite{berridge_am_1997,berridge_calcium_2006,bootman_update_2009}, however, mechanisms for information encoding in the cumulative signal (i.e. area under the curve (AUC), proportional-integral-derivative (PID) controller, or duty cycle (DC)) have also been proposed \cite{purvis_encoding_2013,uzhachenko_computational_2016,hannanta-anan_optogenetic_2016}. Determining which method of information encoding is relevant to a specific signalling pathway requires determining what type of signal encoding the system is capable of, and whether the downstream effector of the signal is capable of temporal signal integration, high or low pass filtering, or threshold filtering. 

In cardiac myocytes, discrete encoding of multiple Ca$^{2+}$-mediated signals is particularly pertinent because of the essential and continuous role Ca$^{2+}$ plays in excitation-contraction coupling (ECC). Of particular significance is the involvement of Ca$^{2+}$ in hypertrophic growth signaling. How Ca$^{2+}$ can communicate a signal in the hypertrophic signalling pathway concurrent with the cytosolic Ca$^{2+}$ fluxes that drive cardiac muscle contraction is still largely unresolved \cite{roderick_calcium_2007,hohendanner_cytosolic_2015}. Understanding  this mechanism is important as \mynotes{pathological} hypertrophic remodelling is a precursor of heart failure and a common final pathway of cardiovascular diseases including hypertension and coronary disease \cite{jefferies_mechanisms_2018,tham_pathophysiology_2015,gilbert_calcium_2019}. 

%In cardiac myocytes, the frequency of calcium oscillation is controlled by the pacemaker cells. While frequency may still play a role in cardiac hypertrophy, there is strong evidence that \ipr-mediated modification of intracellular calcium dynamics can lead to hypertrophy independent of frequency \cite{bootman2009}.

\begin{figure}[!htp]
\centering
\includegraphics[scale=0.8]{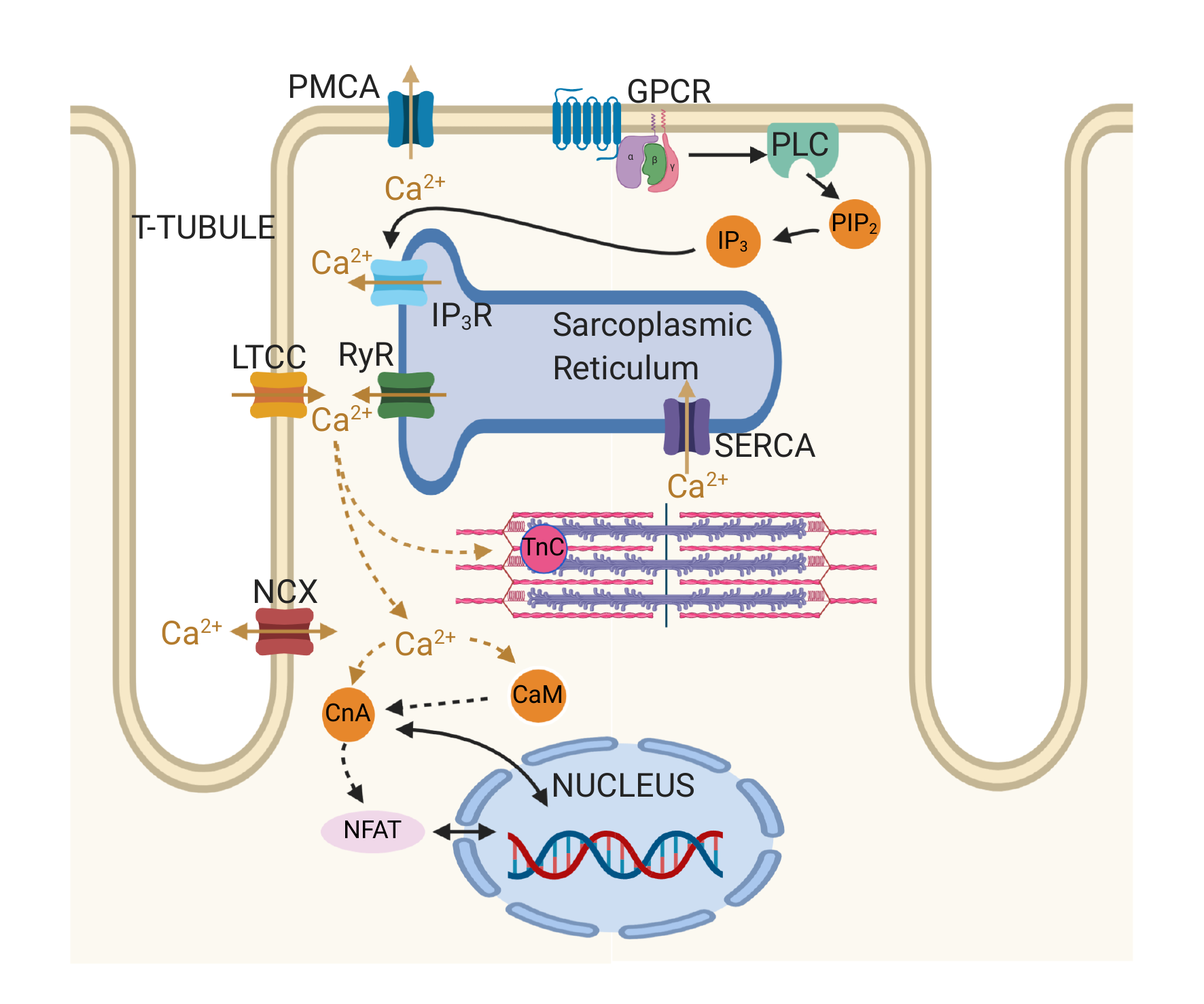}
\caption{Schematic  showing key \cans signalling pathways in the cardiomyocyte. ECC  processes include ryanodine receptors (RyR), L-type \ca channel (LTCCs), sarco-endoplasmic reticulum \ca ATP-ase (SERCA), sodium calcium exchanger (NCX), sarcolemmal calcium pump (PMCA) and Troponin-C (TnC). Growth-related \ipns -- CnA/NFAT signalling processes include inositol 1,4,5-trisphosphate receptors (\iprns), G protein-coupled receptor (GPCR), phospholipase C (PLC), phosphatidylinositol 4,5-bisphosphate (PIP\textsubscript{2}), calmodulin (CaM), calcineurin (CnA) and nuclear factor of activated T-cells (NFAT).}
\label{fig:pathways}
\end{figure}

During each heartbeat, on depolarisation of the membrane Ca$^{2+}$ enters the cell via L-type Ca$^{2+}$ channels (LTCC), triggering larger Ca$^{2+}$ release from the sarcoplasmic reticulum (SR) via ryanodine receptors (RyRs), which then induces contraction. The activation of Ca$^{2+}$  release via RyRs by the Ca$^{2+}$ arising via LTCCs is known as calcium-induced calcium release (CICR), and results in a 10-fold increase in cytosolic Ca$^{2+}$ concentration (relative to resting Ca$^{2+}$ concentration of $\sim$100 nM). Sarco-endoplasmic reticulum Ca$^{2+}$ pumps (SERCA) and other Ca$^{2+}$ sequestration mechanisms subsequently withdraw the released Ca$^{2+}$ back into the SR and out of the cytosol \cite{hinch_simplified_2004,vierheller_multiscale_2015} reverting the cell to its relaxed state. Ca$^{2+}$ also plays a central role in hypertrophic signalling. Hypertrophic stimuli such as endothelin-1 (ET-1) bind to G-protein-coupled receptors at the cell membrane to stimulate generation of the intracellular signalling molecule inositol 1,4,5-trisphosphate (\ipns). After \ip binds to and activates its cognate receptor, inositol 1,4,5-trisphosphate receptors (\iprns), on the SR and nuclear envelope, Ca$^{2+}$ is released into the cytosol and nucleus respectively \cite{higazi_endothelin-1-stimulated_2009,harzheim_increased_2009} (see Figure \ref{fig:pathways}). \mynotes{This \ca signal arising from \iprs has been shown in multiple mammalian species to produce a distinct \ca signal that, through activation of pro-hypertrophic pathways including those involving NFAT, induces hypertrophy within cardiomyocytes. \cite{higazi_endothelin-1-stimulated_2009,nakayama_ip3_2010,rinne_isoform-_2010}.}

In healthy adult rat ventricular myocytes (ARVMs), various effects of \ip on global \ca transients associated with ECC have been described, summarised in Table \ref{tab:exp}. While application of GPCR agonists that stimulate \ip generation produces robust effects on ECC associated \ip transients and contraction, the \mynotes{direct} contribution of \ip to these actions varies between studies \cite{signore_inositol_2013,proven_inositol_2006,harzheim_increased_2009,domeier_ip3_2008,ljubojevic_early_2014,olivares-florez_nuclear_2018}.  For example, in rabbit the effect of ET-1 on \ca transient amplitude is sensitive to the \ipr inhibitor 2-APB \cite{domeier_ip3_2008}, whereas in healthy rats \ipr inhibition with 2-APB was without effect \cite{smyrnias_contractile_2018}. In mice 2-APB abrogated an increase in ECC associated \ca transients brought about by AngII \cite{olivares-florez_nuclear_2018}. Responses have also been variable when \ip was directly applied to cardiac myocytes. In healthy rat, \ip produced no or a modest effect on \ca transient amplitude \cite{proven_inositol_2006,harzheim_increased_2009}, whereas in rabbit \cite{domeier_ip3_2008} a more substantial effect was observed. These differences in the effect of \ip have been ascribed in part to the greater dependence of rat myocytes on SR \ca release to the \ca transient than rabbit myocytes\cite{domeier_ip3_2008}. Notably, both ET-1 and \ip elicit arrhythmogenic effects whereby they promote the generation of \mynotes{spontaneous calcium transients}, manifest as a prolonged \ca transient with additional peaks, and they increase the frequency of \ca sparks \cite{proven_inositol_2006,harzheim_increased_2009,domeier_ip3_2008,nakayama_ip3_2010}. A more profound role for \ip signalling is observed in hypertrophic ventricular myocytes, with ECC-associated \ca transients of greater amplitude reported. Underlying these effects, \ipr expression is elevated in hypertrophy \cite{harzheim_elevated_2010}. Hence, a question remains as to what independent effect \ipr activation has on the cytosolic Ca$^{2+}$ transient in healthy ventricular cardiac myocytes.

\begin{table}[htp]
	\centering
%	\textbf{Experimental data on hypertrophy and \ip} \\
%	\vspace{6pt}
%	\rowcolors{2}{White}{LightBlue}
%
    {\footnotesize
	\begin{tabular}{p{2.5cm} p{2.5cm} p{2.5cm} p{2.5cm}} 
	\hline \rule{0pt}{3ex}    
	\textbf{Cell State} & & \textbf{\ip} & \textbf{ET-1}\vspace{3pt} \\ \hline \rule{0pt}{4ex}  
	Rat 
	    & \pbox{2.5cm}{Amplitude: \\ Duration: \\Basal \cans: \\SCTs:}
		& \pbox{2.5cm}{ 
		r\ding{115}\textsuperscript{\cite{proven_inositol_2006}} \, r\ding{117}\textsuperscript{\cite{harzheim_increased_2009}}\\ \textbf{--}\\ r\ding{117}\textsuperscript{\cite{harzheim_increased_2009}}\\ r\ding{115}\textsuperscript{\cite{proven_inositol_2006}} \, r\ding{115}\textsuperscript{\cite{harzheim_increased_2009}}} 
		& \pbox{3.8cm}{ 
		r\ding{115}\textsuperscript{\cite{proven_inositol_2006}} \, r\ding{117}\textsuperscript{\cite{higazi_endothelin-1-stimulated_2009}} \, r\ding{115}\textsuperscript{\cite{harzheim_increased_2009}}\\ \textbf{--}\\ r\ding{117}\textsuperscript{\cite{harzheim_increased_2009}}\\ r\ding{115}\textsuperscript{\cite{proven_inositol_2006}} \, r\ding{115}\textsuperscript{\cite{harzheim_increased_2009}}} \vspace{1pt}\\ \hline
	Other species
		& \pbox{2.5cm}{Amplitude: \\ Duration: \\Basal \cans: \\SCTs:}
		& \pbox{2.5cm}{ 
		{m\ding{115}}\textsuperscript{\cite{signore_inositol_2013}} \, {m\ding{117}}\textsuperscript{\cite{escobar_role_2012}} \\ \textbf{--}\\ {m\ding{115}}\textsuperscript{\cite{escobar_role_2012}}\\
		\textbf{--}}
		& \pbox{2.5cm}{ {h\ding{115}}\textsuperscript{\cite{signore_inositol_2013}} \, {m\ding{115}}\textsuperscript{\cite{signore_inositol_2013}} \\
		\textbf{--}\\
		{m\ding{115}}\textsuperscript{\cite{signore_inositol_2013}} \, {b\ding{115}}\textsuperscript{\cite{domeier_ip3_2008}}\\ {h\ding{115}}\textsuperscript{\cite{signore_inositol_2013}} \, {m\ding{115}}\textsuperscript{\cite{signore_inositol_2013}}}
		\vspace{1pt}\\ \hline  \rule{0pt}{3ex}    
	\end{tabular}
	}
	\caption{Summary of experimentally observed changes to the Ca$^{2+}$ transient in normal healthy ventricular myocytes in rat and other species following addition of \ip and ET-1. SCTs: spontaneous Ca$^{2+}$ transients; \ding{115} indicates an increase; \ding{116} a decrease; \ding{117} indicates no significant change reported; r indicates rat, b indicates rabbit, h indicates human, m indicates mouse; dashes indicate no data found. The model developed in this work is primarily parameterised with rat data.}
	\label{tab:exp}
\end{table}

%\ca release via \iprs has been experimentally shown to increases the amplitude of the ECC calcium transient in spontaneously hypertensive rats \cite{harzheim_increased_2009}. This is consistent with the observation that these hypertrophic hearts also exhibit increased \ipr expression. Pre-hypertrophic cardiac cells, however, have either been observed to exhibit no change or only moderate change (~1.2 fold change) in amplitude of the calcium transient soon after hypertrophic stimulation \cite{higazi_endothelin-1-stimulated_2009}. The amplitude of the calcium transient progressively increased over a 25 minute period when cells were superfused with ET-1 \cite{proven_inositol_2006}. But this was also accompanied by an increase in frequency of spontaneous diastolic calcium transients, which may have been triggered by ET-1 independent of \ipr \ca release \cite{higazi_endothelin-1-stimulated_2009}. 

The individual behaviour of \ipr channels and their dependence on Ca$^{2+}$, \ipns, and ATP in cardiac and other cell types has been explored in a number of studies \cite{foskett_inositol_2007,ramos-franco_single-channel_2000,siekmann_statistical_2014,siekmann_data-driven_2019}. These studies have formed the basis of several computational models of \ipr type I isoforms \cite{siekmann_statistical_2014,cao_deterministic_2014,sneyd_dynamical_2017} fitted to stochastic single-channel data \cite{siekmann_mcmc_2012}. However, properties of \ipr channel activity within the cardiomyocyte, such as gating state transition rates and their dependency on \ip and \cans, have not been directly measured. 
In this study we have taken the experimental studies on rat ventricular cardiomyocytes as a reference point for the observed effects of \ipr activation on cellular Ca$^{2+}$ dynamics and  extended a well-established model of beat-to-beat cytosolic Ca$^{2+}$ transients in rat cardiac cells \cite{hinch_simplified_2004,terkildsen_using_2008} to include a model of type II \ipr\cite{sneyd_dynamical_2017} channels. This deterministic, compartmental model of ECC enables us to investigate biophysically plausible mechanisms by which \ipr activation could affect Ca$^{2+}$ dynamics at the whole cell scale, while avoiding the computational complexity associated with detailed stochastic and spatial modelling. Specifically, it enables us to explore the parameter ranges of \iprns-mediated \ca release that modify the global cytosolic \ca transient to encode information for hypertrophic signalling to the nucleus.  

A number of transcription factors transduce changes in \ca to activate hypertrophic gene transcription. Of particular note is  Nuclear Factor of Activated T-cells (NFAT). There are five known NFAT isoforms expressed in mammals, four of these are found in cardiac cells \cite{wilkins_targeted_2002,rinne_isoform-_2010}. To initiate hypertrophic remodelling, the hypertrophic Ca$^{2+}$ signal, in conjunction with calmodulin (CaM) and calcineurin (CnA) leads to dephosphorylation of cytosolic NFAT. Upon dephosphorylation NFAT translocates to the nucleus where, in coordination with other proteins, it activates expression of genes responsible for hypertrophy \cite{molkentin_calcineurin-dependent_1998}. Several studies have focused on characterising the Ca$^{2+}$ dynamics necessary to activate NFAT and initiate hypertrophy \mynotes{\cite{tomida_nfat_2003,colella_ca2+_2008,saucerman2008,rinne_isoform-_2010,ulrich2012,yissachar_dynamic_2013,hannanta-anan_optogenetic_2016,kar_control_2016}} and have shown NFAT to be a Ca$^{2+}$ signal integrator \cite{tomida_nfat_2003}. Furthermore, a recent study by \citet{hannanta-anan_optogenetic_2016} used direct optogenetic control of cytosolic Ca$^{2+}$ transients in HeLa cells to demonstrate that the transcriptional activity of NFAT4 (also known as NFATc3), a necessary NFAT isoform in the hypertrophic pathway \cite{wilkins_targeted_2002}, can be up-regulated by increasing the residence time of Ca$^{2+}$ in the cytosol within each oscillation. The increased residence time of Ca$^{2+}$, referred to as the `duty cycle', is the ratio between the area under the Ca$^{2+}$ transient curve divided by the maximum possible area, as calculated by the product of transient amplitude and period (see Figure \ref{fig:cartoontrans}A). \mynotes{The \ca duty cycle is therefore distinct from the average \ca concentration. \citet{hannanta-anan_optogenetic_2016}} showed that increasing the duty cycle had a proportionally greater effect on NFAT transcriptional activity than changing either the frequency or amplitude of the cytosolic Ca$^{2+}$ oscillations. This suggests an increased \ca duty cycle as a possible mechanism by which Ca$^{2+}$ release through \ipr channels can effect hypertrophic signaling. 

Here, using a mathematical model of beat-to-beat cytosolic Ca$^{2+}$ transients \mynotes{in rat ventricular myocytes}, coupled to \ipr channel \ca release, we show that \ipr activation in the cytosol can increase the duty cycle of the cytosolic Ca$^{2+}$ transient. 
%thus providing a plausible mechanism for activation of NFAT.
%in the cardiac cell. 
% Other important modelling work on IP3R type 1 incl. young1992,swillens1999,swaminathan2009
%
We establish model feasibility through parameter sensitivity analysis, which shows that this behaviour does not depend sensitively on model parameter values. Furthermore we identify conditions necessary for \ipr channel activation to alter Ca$^{2+}$ transient amplitude, width, basal \ca and duty cycle, as identified in different experimental studies, and compare model simulations to published experimental data summarised in Table \ref{tab:exp}. \mynotes{Finally, we couple simulations of cytosolic \ca dynamics to a model of downstream CaM/CnA/NFAT activation and show that the duty cycle of the \ca transient highly correlates with  the activated nuclear NFAT (the proportion of NFAT which is dephosphorylated and translocated to the nucleus).}
\mynotes{These findings suggest \ipr activity can increase the cytosolic \ca duty cycle, thus providing a mechanism for \ipns-dependent activation of NFAT for hypertrophic signalling in the cardiomyocyte.}

\section*{Methods}
We developed a computational model of RyR- and \iprns-mediated \ca fluxes in the adult rat ventricular myocyte. Model simulations were performed using the ode15s ODE solver from MATLAB 2017b (The MathWorks Inc., Natick, Massachussetts) with relative and absolute tolerances $1\times10^{-3}$ and $1\times10^{-6}$ respectively. The model equations were simulated at 1 Hz, the original pacing frequency of the \citet{hinch_simplified_2004} model and at 0.3 Hz because it is another common pacing frequency in experimental studies of \ip and Ca$^{2+}$ in cardiomyocytes \cite{proven_inositol_2006,harzheim_increased_2009}. The model was paced until the normalised root mean square deviation (NRMSD) between each subsequent beat was below $1\times10^{-3}$, and all but the last oscillation discarded to eliminate transient behaviours (see Figure \ref{fig:cartoontrans}B). Initial conditions were set to the basal \ca level of the model at dynamic equilibrium with inactive \ipr channels, determined after running the base model until the NRMSD was also below $1\times10^{-3}$. % All model equations and parameter values were taken from \citet{hinch_simplified_2004}, other than those described below relating to \ipr channels and the fluorescent dye.

\subsection*{Model Equations}
The compartmental model of rat left ventricular cardiac myocyte Ca$^{2+}$ dynamics is based on the \citet{hinch_simplified_2004} model of ECC, with the addition of \ipr Ca$^{2+}$ release modelled using the Siekmann-Cao-Sneyd model \cite{sneyd_dynamical_2017}. The Hinch model is an established whole cell model of rat cardiac Ca$^{2+}$ dynamics that describes the flux through the major Ca$^{2+}$ channels and pumps on the cell and SR membranes and the effects of applying a voltage across the cell membrane. The parameters for the Hinch component of our model were maintained from the original except for those of the driving voltage. This was shortened to better approximate the rat action potential \cite{pandit_mathematical_2003} (see Figure S1).
The Ca$^{2+}$ in the cytosol is governed by the following ODE:
\begin{align} 
\deriv{[{\rm Ca}^{2+}]_{\rm cyt}}{t} &= \beta_{\rm fluo} \cdot \beta_{\rm CaM} \cdot \left(I_{\rm CaL} + I_{\rm RyR} - I_{\rm SERCA} + I_{\rm IP\textsubscript{3}R} + I_{\rm other}\right)\\
	I_{\rm other} &= I_{\rm SRl} + I_{\rm NCX} - I_{\rm PMCA} + I_{\rm CaB} + I_{\rm TnC}
	\end{align}
A small Ca$^{2+}$ flux through the LTCCs, $I_{\rm CaL}$, activates RyR channels to release Ca$^{2+}$ from the SR into the cytosol at a rate of $I_{\rm RyR}$. Ca$^{2+}$ is resequestered into the SR by SERCA at a rate $I_{\rm SERCA}$. $\beta_{\rm fluo}$ is the rapid buffer coefficient \cite{wagner_effects_1994} for the fluorescent dye in the cytosol and $\beta_{\rm CaM}$ is the rapid buffer coefficient for calmodulin in the cytosol. $I_{\rm other}$ includes Ca$^{2+}$ fluxes such as exchange with the extracellular environment through the sodium-calcium exchanger, $I_{\rm NCX}$; sarcolemmal Ca$^{2+}$-ATPase, $I_{\rm PMCA}$; and the background leak current, $I_{\rm CaB}$; as well as the SR leak current, $I_{\rm SRl}$; and buffering on troponin C, $I_{\rm TnC}$. These fluxes are defined in the SI (section 1).

When the simulation is run with \ip present, there is additionally a flux through the \iprs:
\begin{align}
	I_{\rm IP\textsubscript{3}R} &= {k_{\rm f}\cdot N_{\rm IP\textsubscript{3}R} \cdot  P_{\rm IP\textsubscript{3}R} \cdot \left([{\rm Ca}^{2+}]_{\rm SR} - [{\rm Ca}^{2+}]_{\rm cyt} \right)} \Big/ {V_{\rm myo}}
	\end{align}	
Here $V_{\rm myo}$ is the volume of the cell. $k_{\rm f}$ is the maximum total flux through each \ipr channel; this was chosen to be 0.45 $\mu$m$^3$ms$^{-1}$ unless otherwise stated to create a measurable effect on \ipr channel activation while maintaining plausible total flux. $N_{\rm IP\textsubscript{3}R}$ is the number of \ipr channels in the cell, this was set to 1/50th of the number of RyR channels \cite{moschella_inositol_1993}. \mynotes{We studied the effect of varying $k_{\rm f}$ on \ipns-induced changes to the cytosolic \ca transient in normal cardiomyocytes. Evidently, varying $N_{\rm IP\textsubscript{3}R}$ and varying $k_{\rm f}$ have the same effect on simulated calcium dynamics. While $N_{\rm IP\textsubscript{3}R}$ is known to increase significantly in disease conditions, we have not emphasised it in this study due to our focus on normal cardiomyocytes.} 
 $[{\rm Ca}^{2+}]_{\rm cyt}$ and $[{\rm Ca}^{2+}]_{\rm SR}$ are the \ca concentrations in the cytosol and SR respectively.

$P_{\rm IP\textsubscript{3}R}$ is the \caconc and [\ipns] dependent open probability of the \ipr channels, and is determined using the Siekmann-Cao-Sneyd model \cite{siekmann_kinetic_2012,cao_deterministic_2014,sneyd_dynamical_2017}, which has an in-built delay in response to changing \ca concentration, along with several parameters governing channel activation and inactivation. This model describes $P_{\rm IP\textsubscript{3}R}$ as: 
\begin{align}
	P_{\rm IP\textsubscript{3}R} = \beta \Big/ \left({\beta+k_\beta\cdot\left(\beta+\alpha\right)}\right)
	\end{align}
where  $k_\beta$ is a transition term derived from single-channel \citet{siekmann_kinetic_2012}, $\beta$ describes the rate of activation and $\alpha$ the rate of inactivation: 
\begin{align}
    \beta&=B\cdot m\cdot h\\
	\alpha&=(1-B)\cdot\left(1-m\cdot h_\infty\right)
\end{align}
where $h$ is time-dependent, and $B$, $m$, and $h/h_\infty$ describe the dependence on \ipns,  the dependence on \ca  and the \cans-dependent delay in \ipr gating, respectively. Expressions for these variables are as follows:
\begin{align}
    B &= [{\rm IP}_3]^2 \Big/ \left(K_p^2+[{\rm IP}_3]^2\right)\\
	m &= {[{\rm Ca}^{2+}]^4} \Big/ \left({K^4_c+[{\rm Ca}^{2+}]^4}\right)\\
    \frac{dh}{dt} &= \left({\left(h_\infty-h\right)\cdot\left(K^4_t+[{\rm Ca}^{2+}]^4\right)}\right) \Big/ \left({t_{max}\cdot K^4_t}\right)\\
    h_\infty &= {K^4_h} \Big/ \left({K^4_h+[{\rm Ca}^{2+}]^4}\right)
\end{align}
Here $K_{\rm c}$ and $K_{\rm h}$ are parameters which determine the \cans-dependence of \ipr channel open probability, while $K_t$ and $t_{max}$ are parameters which affect the delay in \ipr response to cytosolic changes. $K_t$ determines the influence of \caconc on the delay, while $t_{max}$ is a temporal scaling factor. 
 
%\begin{figure}[ht]
%	\centering
%	\includegraphics[scale=0.8]{po.pdf}
%	\caption{The effect of [\ca], [\ip], $K_{\rm c}$, and $K_{\rm h}$ on $P_{IP_3R}$ in the Siekmann-Cao-Sneyd \ipr model \cite{siekmann_kinetic_2012,cao_deterministic_2014,sneyd_dynamical_2017}. The coloured bars on the side of each plot show the approximate proportion of \ipr channels that will open for each set of parameters. This is an approximation based on the model equations that does not take into account the built in delay to changing calcium concentrations. Notice how \iprs do not open at physiological \ca concentrations when $K_{\rm h}$ is low (i.e. 80nM or less). For this model, we used the values $K_{\rm h}=2.2\mu$M, $K_{\rm c}=16\mu$M, and \ip$=10\mu$M.}
%	\label{fig:sneyd}
%\end{figure}

%\subsection*{\mynotes{The effect of \ipr activation on SR calcium leak flux}}
\mynotes{We note that the SR leak flux, $I_{\rm SRl}$, is unchanged from the Hinch model, and would include the effects of diastolic \ipr \ca release at normal \ip levels as that model did not explicity include \iprns. However, in the presence of \ipns, \ipr \ca flux during diastole is several orders of magnitude greater than $I_{\rm SRl}$, which is largely dependent on $\rm [Ca^{2+}]_{SR}$, and hence any discrepancy caused by this will have a negligible effect on overall \ca dynamics within the cell (see also Figure S4).}
%
%\mynotes{Conversely, we have not included the effect of \ipr leak in the presence of %\ip on other components of $I_{\rm SR}$ either. It is possible that increased leak %from \iprs promotes increased \ca leakage via other channels. Depending on the %mechanisms behind the various components of $I_{\rm SR}$, $I_{\rm SR}$ could have a %greater impact on overall \ca dynamics during hypertrophic signalling than we have %accounted for.}

Several experimental studies have investigated \ipr activity across a range of Ca$^{2+}$ concentrations with $1$ $\mu$M \ip \cite{ramos-franco_isoform-specific_1998,foskett_inositol_2007}. These studies suggest that \ipr channels would be open, with almost constant $P_{\rm IP\textsubscript{3}R}$ over the full range of cytosolic Ca$^{2+}$ concentrations experienced during ECC in the cardiomyocyte. An \iprns-facilitated SR-\ca leak has been reported to amplify systolic concentrations \cite{zima_ca2_2010,blanch_i_salvador_obstruction_2018} as  seen in most published experiments of \ip enhanced Ca$^{2+}$ transients tabulated in Table \ref{tab:exp}. 
Through parameter sensitivity analysis of this model, we show that in order to be consistent with these observations $P_{\rm IP\textsubscript{3}R}$ must be significantly smaller at resting \ca concentrations than at higher concentrations.

\subsection*{\mynotes{Coupling cytosolic \ca and NFAT activation}}
\mynotes{We coupled the calcium model to the  NFAT model developed by \citet{cooling_sensitivity_2009}, which determines the proportion of total cellular NFAT that is dephosphorylated and translocated to the nucleus for a given cytosolic \ca signal. In this study we have used the model parameters estimated from the data in \citet{tomida_nfat_2003} who measured activation of NFAT4 in BHK cells. Full details of the \citet{cooling_sensitivity_2009} model are given in the Supplementary Information.}

\section*{Results}
\begin{figure}[!htp]
	\centering
	A\raisebox{-0.8\height}{\includegraphics[scale=0.25]{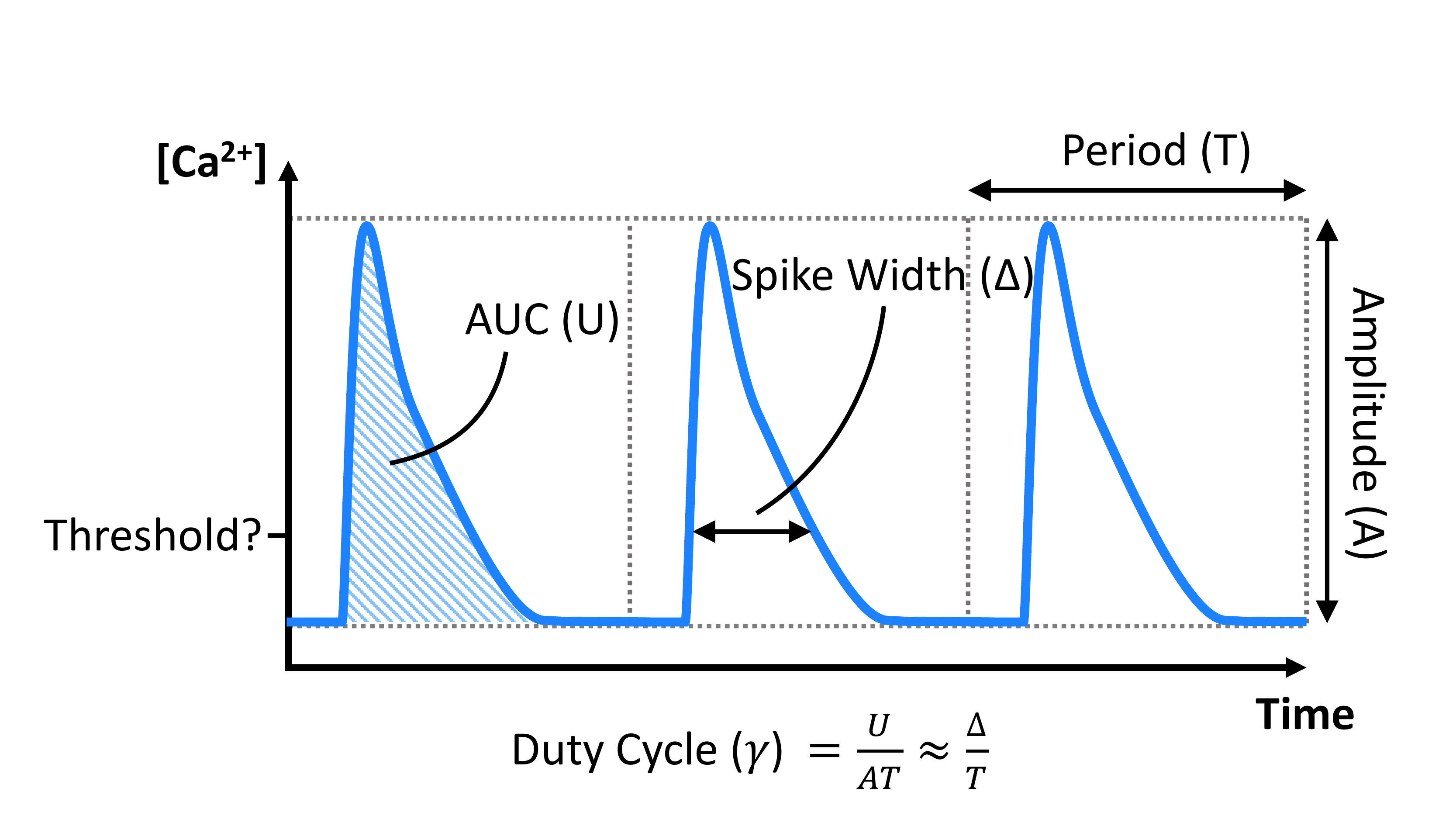}}
	B\raisebox{-0.8\height}{\includegraphics[scale=0.5]{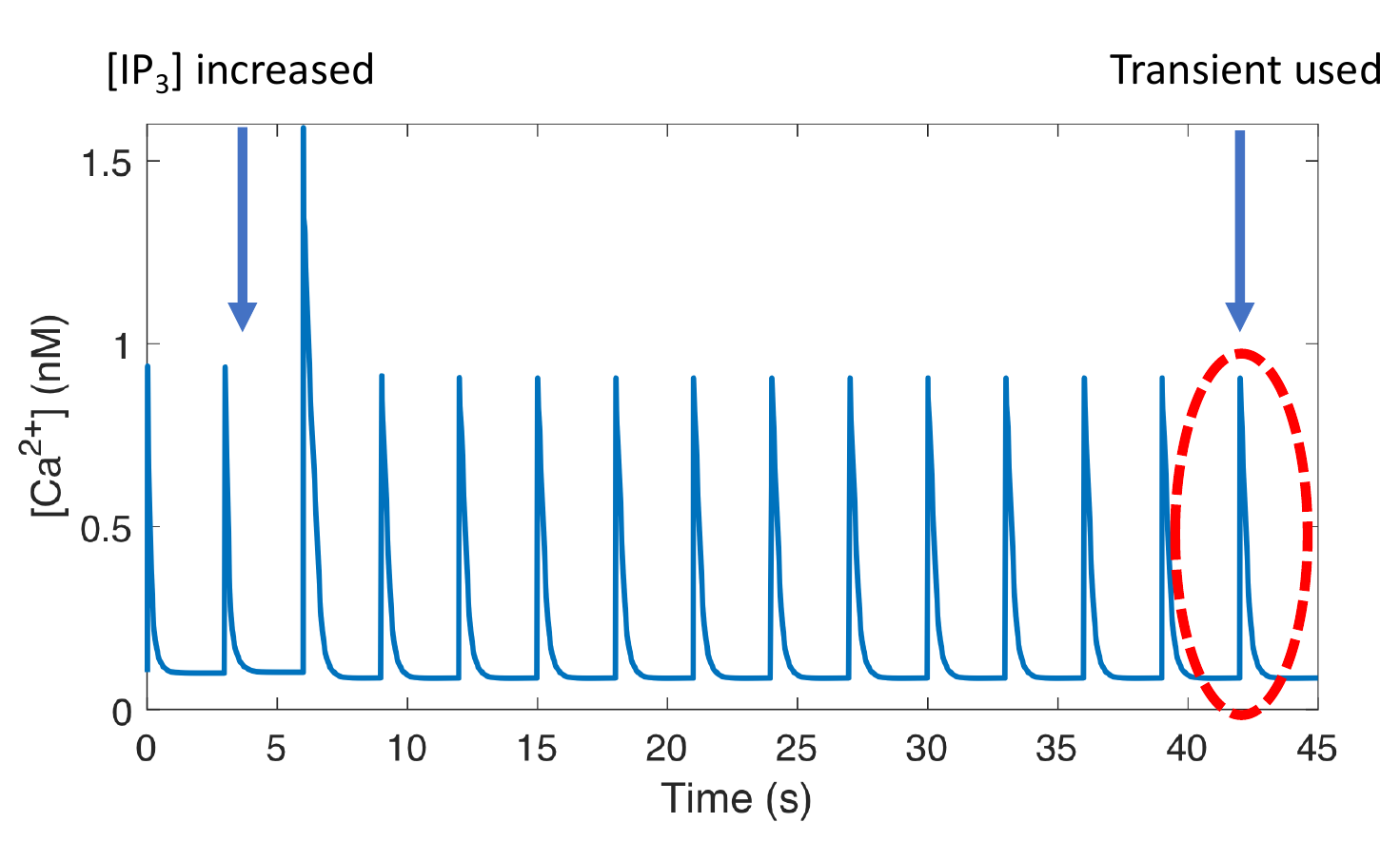}}
	C\raisebox{-0.8\height}{\includegraphics[width=0.9\textwidth]{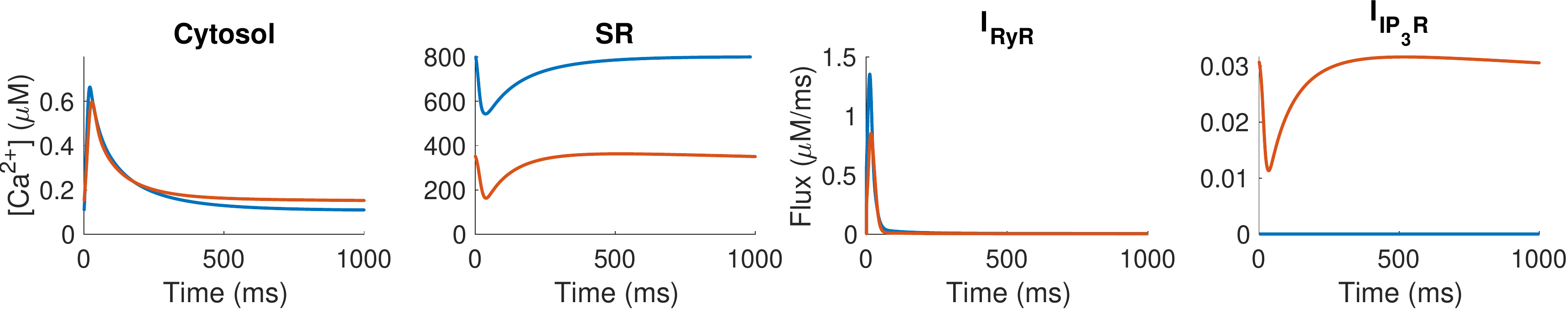}}
	\caption{(A)  The duty cycle, a function of AUC, amplitude, and period, for the cytosolic Ca$^{2+}$ transient. (B) Example of Ca$^{2+}$ transients generated by the model. \mynotes{(C) \ca concentration in cytosol and SR, RyR flux, and \ipr flux in the model with elevated \ip (red) and without \ip (blue). Here \ipr parameters used are taken from \citet{sneyd_dynamical_2017}, with maximum \ipr flux $k_{\rm f}=0.003$ $\mu$m$^3$ ms$^{-1}$.}}
	\label{fig:cartoontrans}
\end{figure}

\mynotes{An example of the model output when run with the original \ipr channel parameter values determined by \citet{sneyd_dynamical_2017} for type I \ipr channels is shown in Figure \ref{fig:cartoontrans}C. Measurements of the properties of \ipr channel activity and their dependence on \ca within cardiomyocytes are sparse in the literature. Therefore we performed a parameter sensitivity analysis by running model simulations over a variety of parameter ranges to explore the dependence of features of the cytosolic calcium transient to \ipr channel parameters.}

\subsection*{Parameter sensitivity analysis}

We conducted a parameter sensitivity analysis to determine the critical parameters related to \ipr activation that affect the shape of beat-to-beat cytosolic Ca$^{2+}$ transients. We used the Jansen method \cite{jansen_analysis_1999} as described in \citet{saltelli_variance_2010} (and summarised in the SI) to calculate the `main effect' and `total effect' coefficients of each of the parameters associated with \ipr channel gating in relation to changes in transient amplitude, full duration at half maximum (FDHM), diastolic \ca and duty cycle (see Table \ref{tab:var_effects}). \citet{saltelli_variance_2010} describe the main effect coefficient as `the expected reduction in variance that would be obtained if [the parameter] could be fixed' and the total effect coefficient as `the expected variance that would be left if all factors but [the parameter] could be fixed', both normalised by the total variance. Both coefficients are included here to provide a complete picture of the impact of each parameter. Simulation parameter values were generated using the MATLAB sobolset function with leap $1\times10^3$ and skip $1\times10^2$.\\

\begin{table}[ht!]
	\centering
	\textbf{Variance-based parameter sensitivity analysis} \\
	\vspace{6pt}
%	\rowcolors{2}{White}{LightBlue}
    {\footnotesize
	\begin{tabular}{p{3.5cm} p{1.5cm} p{1.5cm} p{1.5cm} p{1.5cm} p{1.5cm} p{1.5cm}} 
	    \hline \rule{0pt}{3ex}    
	    \textbf{{Main Effect Coefficients}} &[\ipns] & $t_{max}$ & $K_{\rm c}$ & $K_{\rm h}$ & $K_t$ & $k_{\rm f}$\vspace{2pt} \\ \hline
	    Amplitude   &    \textbf{0.27}   & 0.00 & 0.03 & \textbf{0.19} & 0.00 & 0.03\\
	    FDHM        & \textbf{0.17} & 0.00 & 0.01 & \textbf{0.12} & 0.00 & \textbf{0.50}\\
	    Diastolic \ca & \textbf{0.44} & 0.00 & 0.09 & 0.03 & 0.00 & 0.04\\
	    Duty Cycle  & \textbf{0.23} & 0.00 & 0.01 & \textbf{0.16} & 0.00 & \textbf{0.33}
	    \vspace{1 pt}\\ \hline \rule{0pt}{3ex}    
        \textbf{{Total Effect Coefficients}} &[\ipns] & $t_{max}$ & $K_{\rm c}$ & $K_{\rm h}$ & $K_t$ & $k_{\rm f}$\vspace{2pt} \\ \hline
    	Amplitude   & \textbf{0.63} & 0.04 & \textbf{0.43} & \textbf{0.46} & 0.02 & \textbf{0.13}\\
	    FDHM        & \textbf{0.33} & 0.00 & \textbf{0.19} & \textbf{0.19} & 0.00 & \textbf{0.54}\\
	    Diastolic \ca & \textbf{0.79} & 0.00 & \textbf{0.45} &0.06 & 0.00 & \textbf{0.18}\\
	    Duty Cycle  & \textbf{0.45} & 0.00 & \textbf{0.25} & \textbf{0.24} & 0.00 & \textbf{0.38}
	     \vspace{1 pt}\\ \hline \rule{0pt}{3ex}    
	\end{tabular}
	}
	\caption{Main and total effects of the \ipr gating parameters on Ca$^{2+}$ transient amplitude, duration (FDHM), diastolic Ca$^{2+}$, and duty cycle. \mynotes{Significant values are highlighted in bold font.}}
	\label{tab:var_effects}
\end{table}
% We tuned these parameters to achieve the same result in simulations with hypertrophic levels of \ipr channels. In order for this to happen:
% \begin{itemize}
% 	\item \ipr channels must only open when cytosolic calcium is elevated -- otherwise they drain the SR and reduce the available calcium stored in the SR to the point where it cannot sustain the increase.
% 	\item \ipr channels must be open at the peak of the calcium transient -- to augment the release by RyRs
% 	\end{itemize}
% To achieve this, we used the values $K_{\rm c}=16\mu$M, $K_{\rm h}=2.2\mu$M, $K_t=0.2\mu$M, and $t_{max}=1 \rm{s}^{-1}$ in the subsequent simulations. 
Table \ref{tab:var_effects} shows that the delay parameters $t_{max}$ and $K_t$ do not have a large effect on the cytosolic \ca transient. While they are necessary to describe the effect of \iprns-dominated Ca$^{2+}$ dynamics \cite{sneyd_dynamical_2017}, they contribute only a small amount to the variance. Therefore we decided to fix these parameters in our simulations. 

As expected, the coefficients show that cardiac cell Ca$^{2+}$ dynamics during ECC are most highly sensitive to \ip concentration ([\ipns]) and the maximal flux through each \ipr ($k_{\rm f}$). The maximal flux $k_{\rm f}$ has little effect on transient amplitude, but \mynotes{large} influence on duration and duty cycle; while [\ipns] has the greatest effect on the change in amplitude and diastolic \ca concentration.

\begin{figure}[!htp]
	\centering
	\includegraphics[scale=0.9]{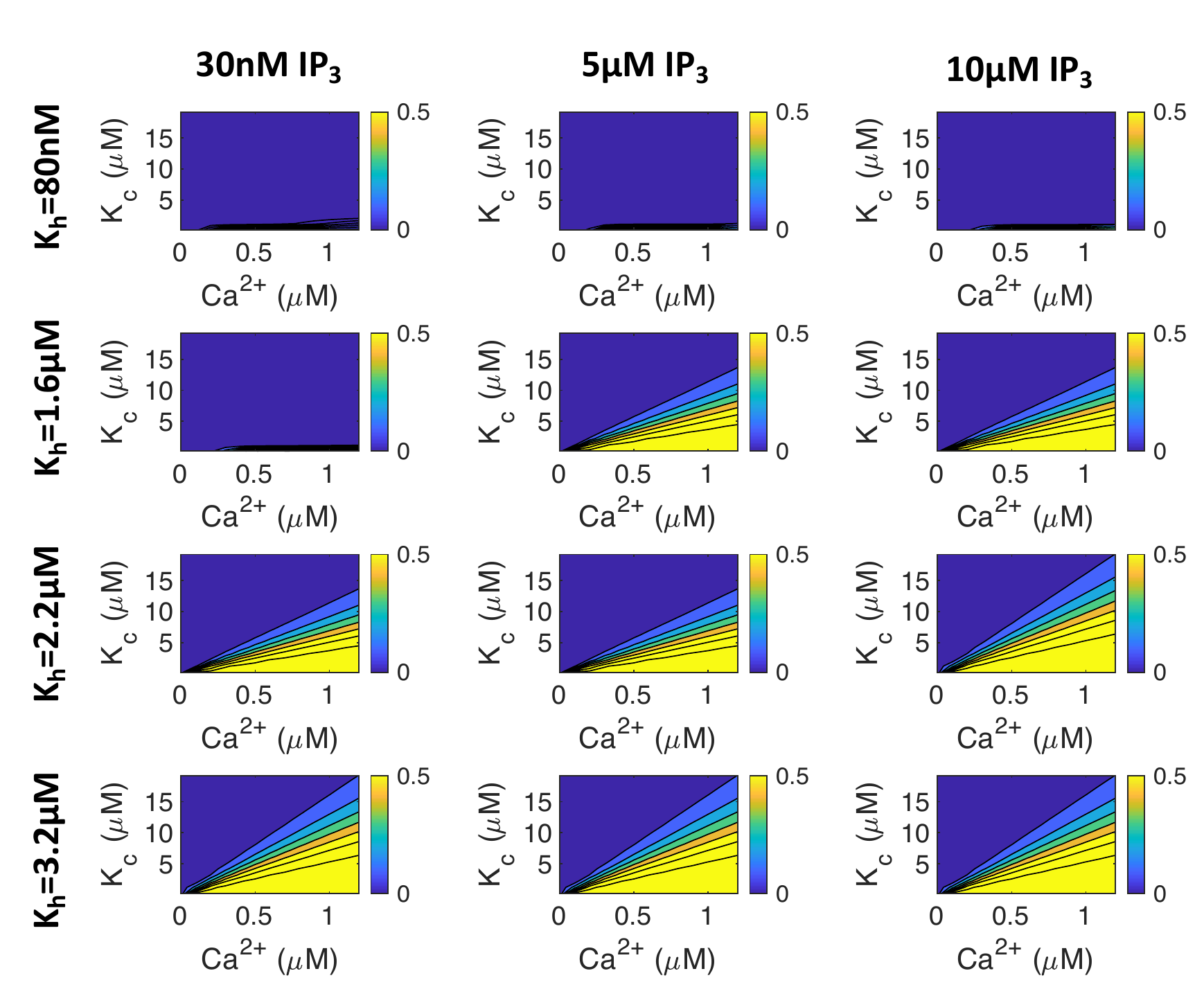}
	\caption{The effect of [\cans], [\ipns], $K_{\rm c}$, and $K_{\rm h}$ on $P_{\rm IP\textsubscript{3}R}$ in the Siekmann-Cao-Sneyd \ipr model \cite{siekmann_kinetic_2012,cao_deterministic_2014,sneyd_dynamical_2017}. The coloured bars on the side of each plot show the proportion of \ipr channels that will open for each set of parameters at steady state. Note that \iprs do not open at physiological \ca concentrations when $K_{\rm h}$ is low (i.e. 80 nM or less). In subsequent simulations we used the value $K_{\rm h}=2.2$ $\mu$M unless otherwise stated.}
	\label{fig:sneyd}
\end{figure}

The gating parameters $K_{\rm c}$ and $K_{\rm h}$ also influence the cytosolic \ca transient. $K_{\rm h}$ affects the [\cans] at which \ipr channels are inhibited and $K_{\rm c}$ affects the [\cans] at which \ipr channels open. We illustrate how these two parameters affect \ipr open probability, $P_{\rm IP\textsubscript{3}R}$, in Figure \ref{fig:sneyd}. Figure \ref{fig:sneyd} also shows how [\ipns]  affects the relationship between $K_{\rm c}$, $K_{\rm h}$, \caconc and $P_{\rm IP\textsubscript{3}R}$. It can be seen that with $K_{\rm h}$ = 80 nM, $P_{\rm IP\textsubscript{3}R}$ will be close to zero regardless of the values of \ca or [\ipns] or $K_{\rm c}$. At $K_{\rm h}=1.6$ $\mu$M and \mbox{[\ipns] $\geq 5$ $\mu$M} $P_{\rm IP\textsubscript{3}R}$ dependence on $K_{\rm c}$ and \ca becomes apparent. Finally, at $K_{\rm h}=3.2$ $\mu$M, $P_{\rm IP\textsubscript{3}R}$ is still dependent on $K_{\rm c}$ and \ca values, but  [\ipns] does not change $P_{\rm IP\textsubscript{3}R}$  significantly. 

From this analysis, we determine that in order for \ipr channels to be active during ECC, $K_{\rm h}$ must be sufficiently high  that \iprs are not inhibited at diastolic \caconcns. Conversely, $K_{\rm c}$ must be low enough that \ipr channels are active at Ca$^{2+}$ concentrations below the systolic \ca peak. Therefore, in the remainder of this study, we fix $K_{\rm h}$ at $2.2$ $\mu$M: high enough to fulfill this condition while low enough that \ipr channels are still affected by [\ipns]. We report simulation results only within the range of $K_{\rm c}$ that exhibits experimentally plausible Ca$^{2+}$ transient properties.

With the plausible range of $K_{\rm h}$ and $K_{\rm c}$ established, we next show the effect of $K_{\rm c}$, $k_{\rm f}$ and [\ipns] on the ECC transient.

\subsection*{\ip concentration and \ipr opening behaviour have the greatest impact on the Ca$^{2+}$ transient}

\begin{figure}[!htp]
	\centering
	\begin{subfigure}[b]{0.45\textwidth}
	%	\centering
		\includegraphics[width=1.\textwidth]{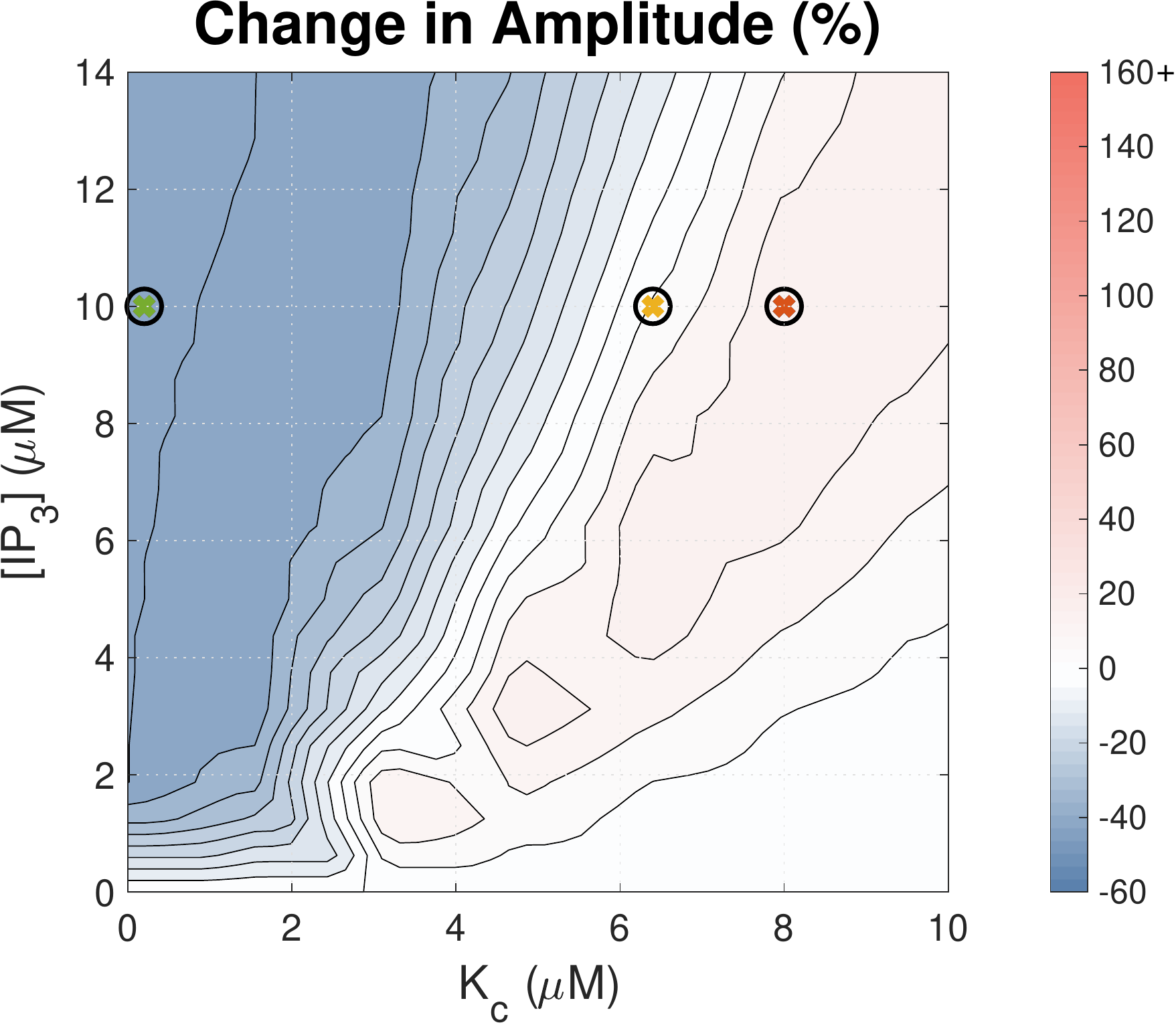}
		\put(-215,200){\Large A}
	\end{subfigure}\hspace{10mm}
	\begin{subfigure}[b]{0.45\textwidth}
	%	\centering
		\includegraphics[width=1.\textwidth]{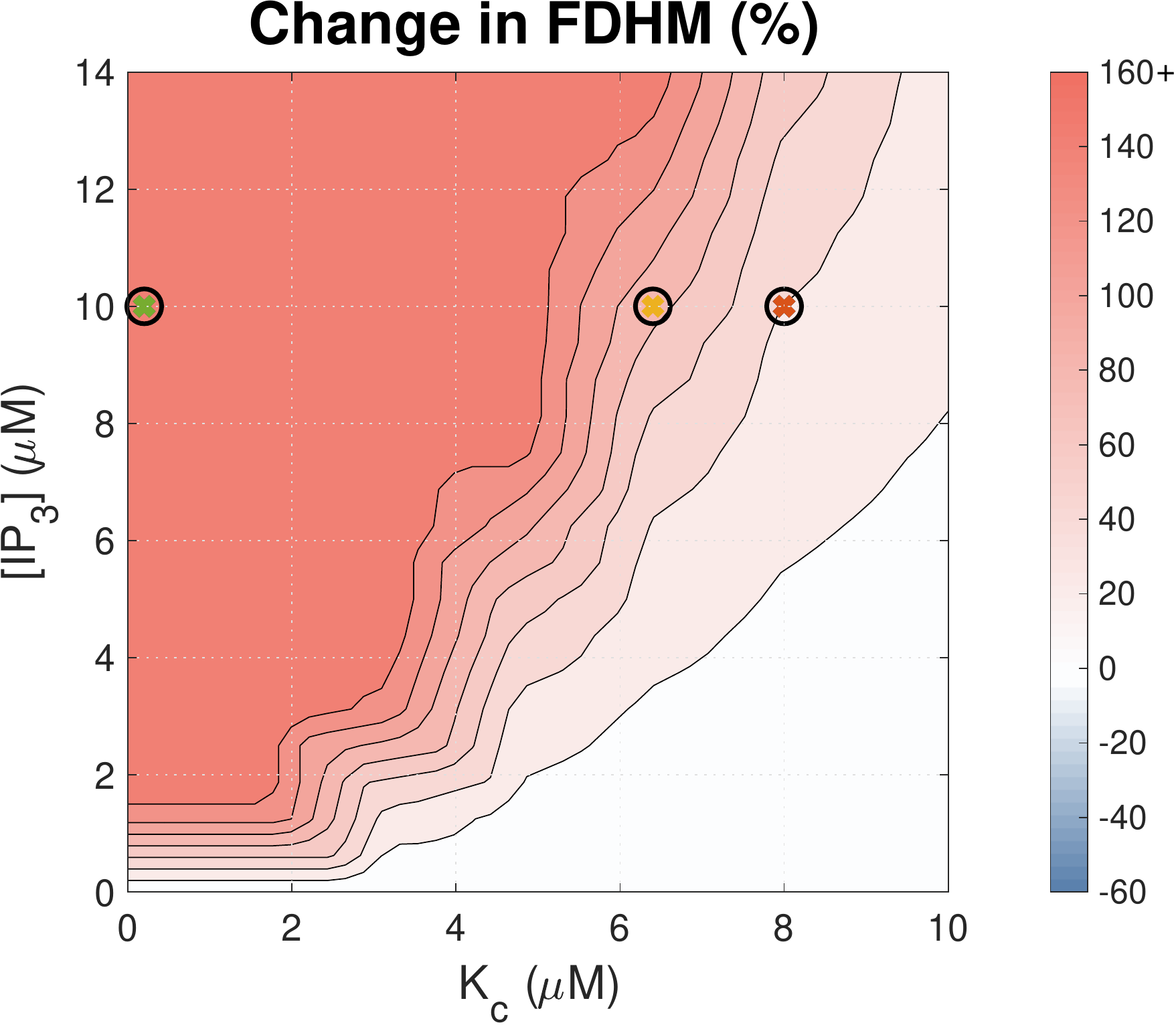}
		\put(-215,200){\Large B}
	\end{subfigure}
	~\\
	\begin{subfigure}[b]{0.45\textwidth}
	%	\centering
		\includegraphics[width=1.\textwidth]{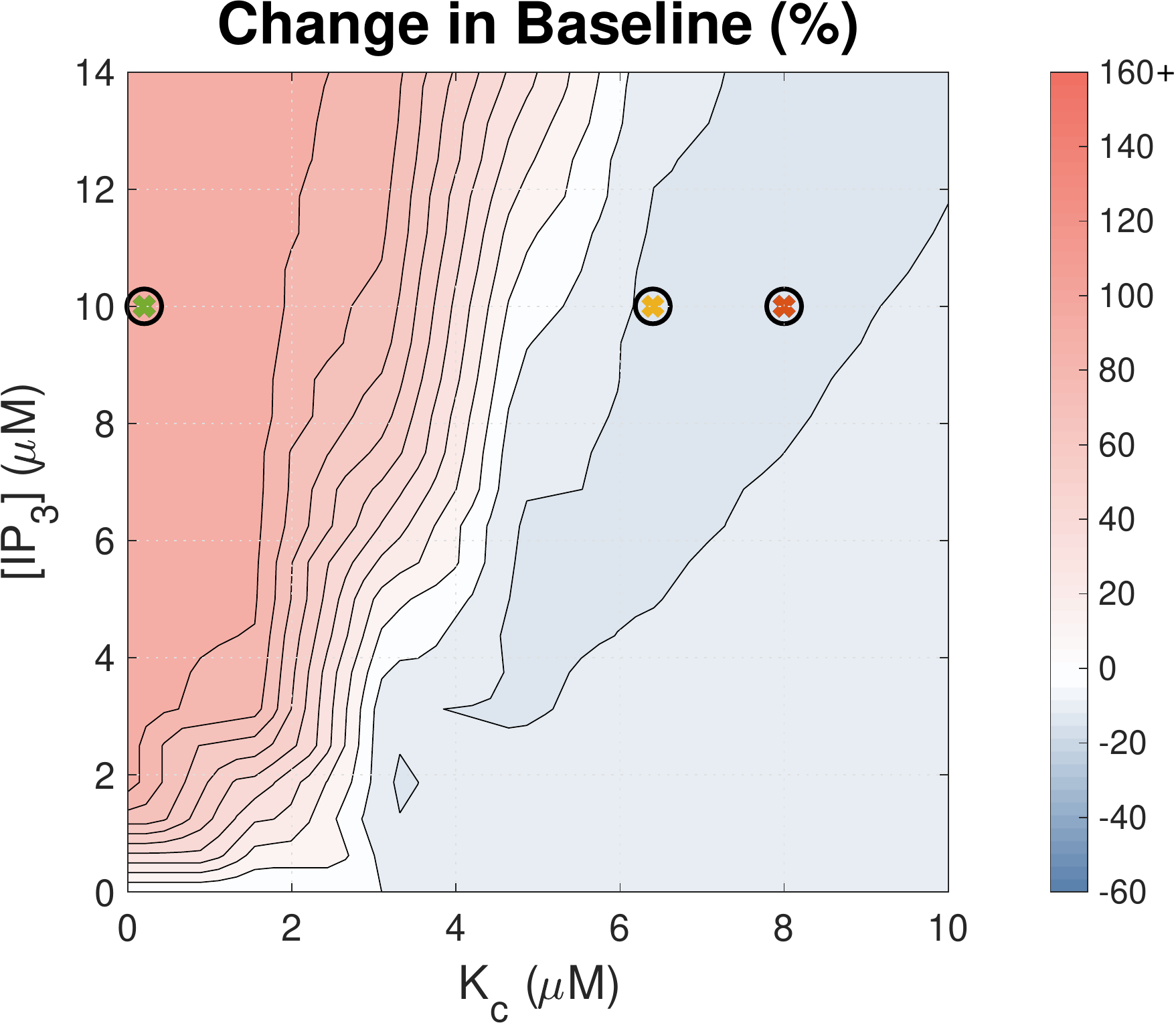}
		\put(-215,200){\Large C}
	\end{subfigure}\hspace{10mm}
	\begin{subfigure}[b]{0.45\textwidth}
	%	\centering
		\includegraphics[width=0.84\textwidth]{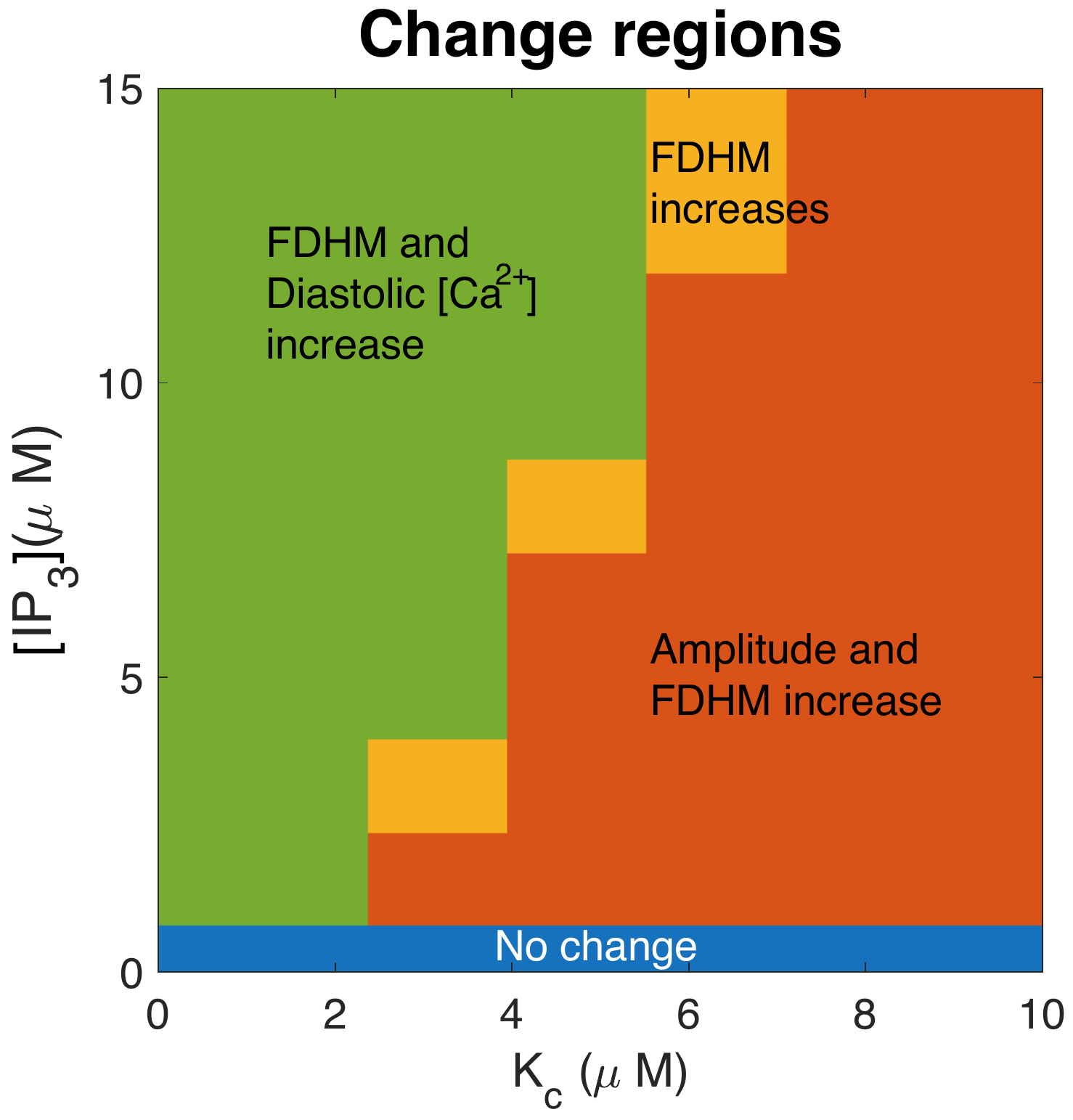}
		\put(-180,200){\Large D}
	\end{subfigure}
	\caption{Effect of \ip concentration and the parameter $K_{\rm c}$ on the Ca$^{2+}$ transient with pacing frequency 1 Hz. These two parameters, along with maximum \ipr flux, $k_{\rm f}$ have the greatest impact when considering the effect of \ipr activation on the Ca$^{2+}$ transient. To better resolve the range in which FDHM changes, all FDHM increases of 45\% and over are shown in the same colour. See Figure \ref{fig:flux_sn} for simulated transients at parameters indicated by crosses. \mynotes{We note that for ease of comparison between figures, in this and in subsequent figures the maximum increase from baseline is cropped at 160\%. Changes greater than this threshold are shown in the same colour.} }
	\label{fig:effect_kc_ip3}
\end{figure}

As summarised in Table \ref{tab:exp}, different experimental studies suggest different effects of \ipr activation on the ECC cytosolic Ca$^{2+}$ transient. Figures \ref{fig:effect_kc_ip3}A-C show quantitative predictions of how much \ca transient properties could be affected by \ipr activation across a range of [\ipns] and \cans-dependent \ipr gating parameter $K_{\rm c}$ values. $k_{\rm f}$ was fixed at 0.45 $\mu$m$^3$ms$^{-1}$ and $K_{\rm h}$ was fixed at 2.2 $\mu$M. 
%The green, orange and red crosses in each of Figure \ref{fig:effect_kc_ip3}A-C mark three examples of types of \ipr defined by $K_{\rm c}$ (and $K_{\rm h}=2.2 \muM$) at [\ipns] of $10 \muM$. Figure \ref{fig:flux_sn} shows corresponding plots of the ECC \ca transient and \ca currents through the \ipr, RyRs, SERCA and NCX for these three sample \ipr types.  

The red region in Figure \ref{fig:effect_kc_ip3}A corresponds to \ipr activation parameters that produce the greatest increase in \ca amplitude. Noteworthy is that the red region depicts moderate  changes in amplitude of $\sim$15\%. This region corresponds to $K_{\rm c}$ values greater than 4 $\mu$M and [\ipns] greater than \mbox{2 $\mu$M}. With $K_{\rm h}$ set at 2.2 $\mu$M, this corresponds to the middle and far-right plots of $P_{\rm IP\textsubscript{3}R}$ in Figure \ref{fig:sneyd}. The middle subfigure shows that with $K_{\rm c}$ greater than 4 $\mu$M \ipr channels would open only at \ca concentrations greater than the diastolic concentration of $\sim$0.1 $\mu$M. The plot also shows that \iprs would remain active at \cans greater than the systolic peak concentration of $\sim$1 $\mu$M \mynotes{\cite{greenstein2002}}. Figure \ref{fig:effect_kc_ip3}B further indicates that the increase in peak amplitude is accompanied by an increase in transient duration (FDHM). However, this change may be small, particularly at \ip concentrations lower than 1 $\mu$M. In Figure \ref{fig:effect_kc_ip3}C it can be seen that the diastolic Ca$^{2+}$ concentration decreases moderately ($\sim$10\%) in the parameter range where the amplitude is maximised (Figure \ref{fig:effect_kc_ip3}A).

Figure \ref{fig:effect_kc_ip3}B shows that FDHM of the Ca$^{2+}$ transient increases whenever \iprs are active. This increase is greater with greater concentrations of \ip and with lower values of $K_{\rm c}$. Figure \ref{fig:effect_kc_ip3}C indicates that $K_{\rm c}$ and [\ipns] have a similar effect on the diastolic Ca$^{2+}$ concentration except that the location of the red and orange cross predicts a small ($\sim$10\%) drop in diastolic \cans. In all three Figures (A-C) there is little change when [\ipns] is low and $K_{\rm c}$ is high (bottom right corner of each image). This is a regime in which the \ipr channels barely open in response to ECC transients. For comparison, Supplementary Figure S2 shows the same simulations as Figure \ref{fig:effect_kc_ip3} at a commonly used experimental pacing frequency of 0.3 Hz, showing similar trends.

%Interestingly, \ipr activation only ever increases the FDHM of the transient (see the scale bar next to Figure \ref{fig:effect_kc_ip3}B). This apparent delay in calcium re-uptake appears due to the continued release through \ipr channels after RyRs have closed \ATcom{(see Figure \ref{fig:flux_sn})}. The initial rapid release into the cytosol through RyRs and \iprs is followed by re-uptake as the SR refills via SERCA. During this recovery, a smaller (in magnitude) period of calcium release (approx. 200ms, see Figure \ref{fig:flux_sn}) is observed from \iprs which contributes to this delay. The slower release through \ipr channels here is a result of a smaller proportion of the channels opening and a decrease in SR calcium stores at this stage.

In order to compare our simulation results with the experimental observations summarised in Table \ref{tab:exp} we divided the parameter space shown in Figures \ref{fig:effect_kc_ip3}A-C into four regions, shown in Figure \ref{fig:effect_kc_ip3}D. In the red region, amplitude and FDHM increase. In the orange region only FDHM increases. In the green region FDHM and diastolic \caconc increase but amplitude decreases. Comparing to the experimental observation of  amplitude increase summarised in Table 1, the red region appears to describe the most plausible parameter range. Figure \ref{fig:effect_kc_ip3}D also shows that there is no parameter set where both amplitude and diastolic \ca concentration increase. Furthermore, there is no region in which transients with increased amplitude and decreased duration are observed, as has been reported in ET-1 treated rat ventricular myocyte experiments \cite{moravec_endothelin_1989}. Finally, with the exception of the blue region in which there is no change, we observe that the FDHM increases in all parameter regimes.

\begin{figure}[!htp]
	\centering
	\includegraphics[width=0.9\textwidth]{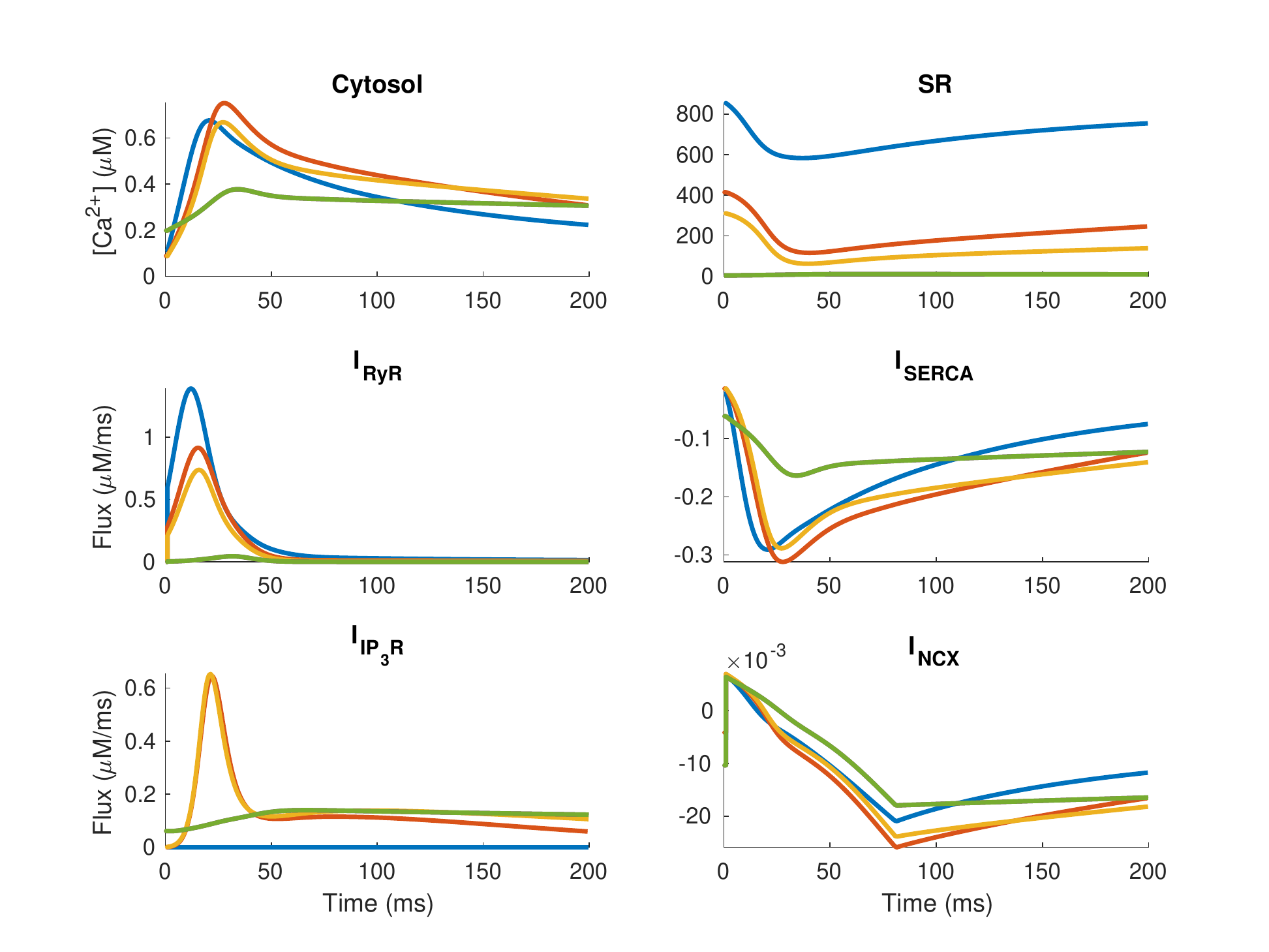}
	\caption{Simulated ECC transient and fluxes in the absence (blue) and presence of \ip, corresponding to low (green), medium (orange) and high (red) values of $K_{\rm c}$. 
	%\ipr gating is based on the Siekmann-Cao-Sneyd model \cite{siekmann_kinetic_2012,cao_deterministic_2014,sneyd_dynamical_2017} with parameters tuned for an increase in systolic calcium. 
	With $K_{\rm c}=8$ $\mu$M (orange), \ipr channels open only at \ca concentrations greater than $0.1$ $\mu$M. This results in increased peak in cytosolic \ca transients and depleted SR \ca stores. Parameters here were selected to show: absence of \ipr channels (blue), increased transient amplitude (orange, red) and \iprs parameterised as described in the original Siekmann-Cao-Sneyd model (green). \ip concentration is $10$ $\mu$M and pacing frequency 1 Hz in all simulations. 
	The sign of $I_{\rm NCX}$ indicates whether Ca$^{2+}$ is moving into (positive) or out of (negative) the cell. }
	\label{fig:flux_sn}
\end{figure}

To examine these results further, we investigated model behaviour in different regions of Figure \ref{fig:effect_kc_ip3}D, shown in Figure \ref{fig:flux_sn} and marked as green, red and orange crosses in Figure \ref{fig:effect_kc_ip3}A-C. Comparing the green cytosolic profiles (corresponding to the green region in Figure \ref{fig:effect_kc_ip3}D) and blue cytosolic \ca profiles (corresponding to no \ipr activation) in Figure \ref{fig:flux_sn}, we find that \ipr opening at diastolic Ca$^{2+}$ levels and \ipr inhibition at Ca$^{2+}$ levels below peak transient concentrations generates a flatter Ca$^{2+}$ transient. This is the result of a gradual depletion of SR Ca$^{2+}$ stores from \iprs opening. This subsequently leads to lower \ca release through RyR and \ipr channels.

Interestingly, a delayed time to peak is observed with \ipr activation in all regimes selected. \mynotes{With the reduction in SR load due to \ipr activation, we find reduced \ca flux through RyRs. In order to maintain or increase \ca transient amplitude after activation, the \ipr channels must compensate for the drop in RyR flux. As the spike in \ipr flux is in response to \ca release from RyR channels, and initial RyR-mediated \ca release is slower with lower SR \ca stores, it delays the time between cell stimulation and \ca transient peak.}

The increase in FDHM of the transient from \ipr activation apparent in Figure \ref{fig:effect_kc_ip3}B can be explained by continued release of \ca through \ipr channels after RyRs have closed in Figure \ref{fig:flux_sn}. The slower release through \ipr channels after RyRs close is a result of a smaller proportion of the channels opening and a decrease in SR Ca$^{2+}$ store load.

%\subsection*{Increasing [\ip] alters \ipr gating and increases the effect of \iprs}
%
%\ATcom{Isn't it just a shorter repetition of things mentioned in previous subsection?\\}
%In our model, the primary effect of increasing \ip concentration is an increase in the \ca release by \ipr channels, increasing transient duration (FDHM) and decreasing SR stores (see Figure \ref{fig:effect_kc_ip3}B and (see Figure \ref{fig:effect_kc_ip3}C). The effect of \ip concentration on transient amplitude and diastolic calcium concentrations is more complicated because [\ip] also increases the calcium concentration required to activate \ipr channels. For the same $K_{\rm c}$ values, an increase in [\ip] results in a decrease in the effect of low \ca concentrations on \ipr activation. To a point, greater \ip concentrations will result in greater \ca transient amplitude and duration. Above a certain concentration, \iprs are active at every point during the calcium transient (see Figure \ref{fig:effect_kc_ip3}A). Sustained \ipr \ca release causes a decrease in SR calcium stores, and hence systolic calcium while increasing transient duration (Figure \ref{fig:effect_kc_ip3} and \ref{fig:sneyd}).

\subsection*{Maximum flux through \iprs can increase signal duration}

\begin{figure}[!htp]
	\centering
	\begin{subfigure}[b]{0.45\textwidth}
		\includegraphics[width=1.\textwidth]{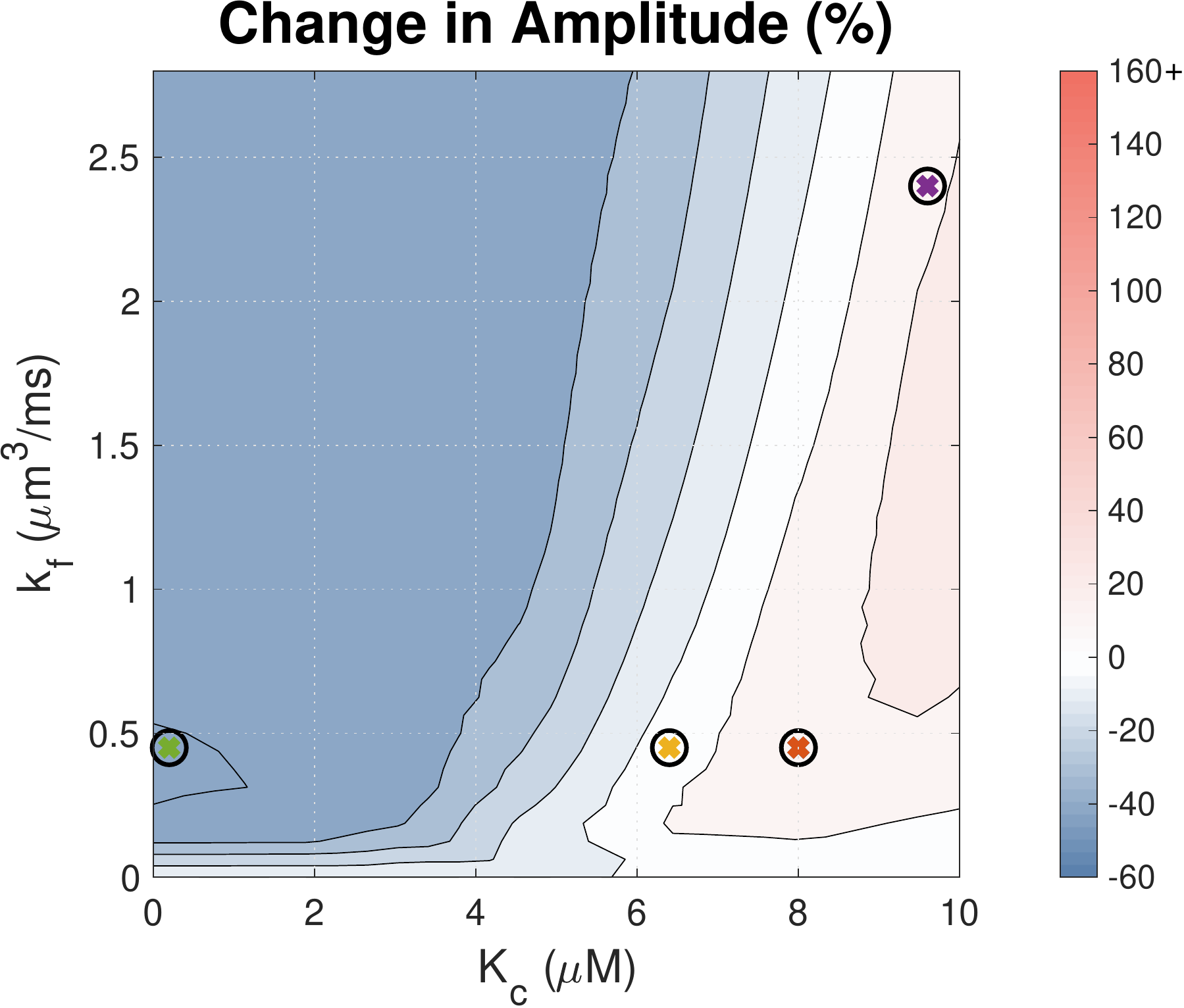}
		\put(-215,200){\Large A}
	\end{subfigure}
	\begin{subfigure}[b]{0.45\textwidth}\hspace{5mm}
		\includegraphics[width=1.\textwidth]{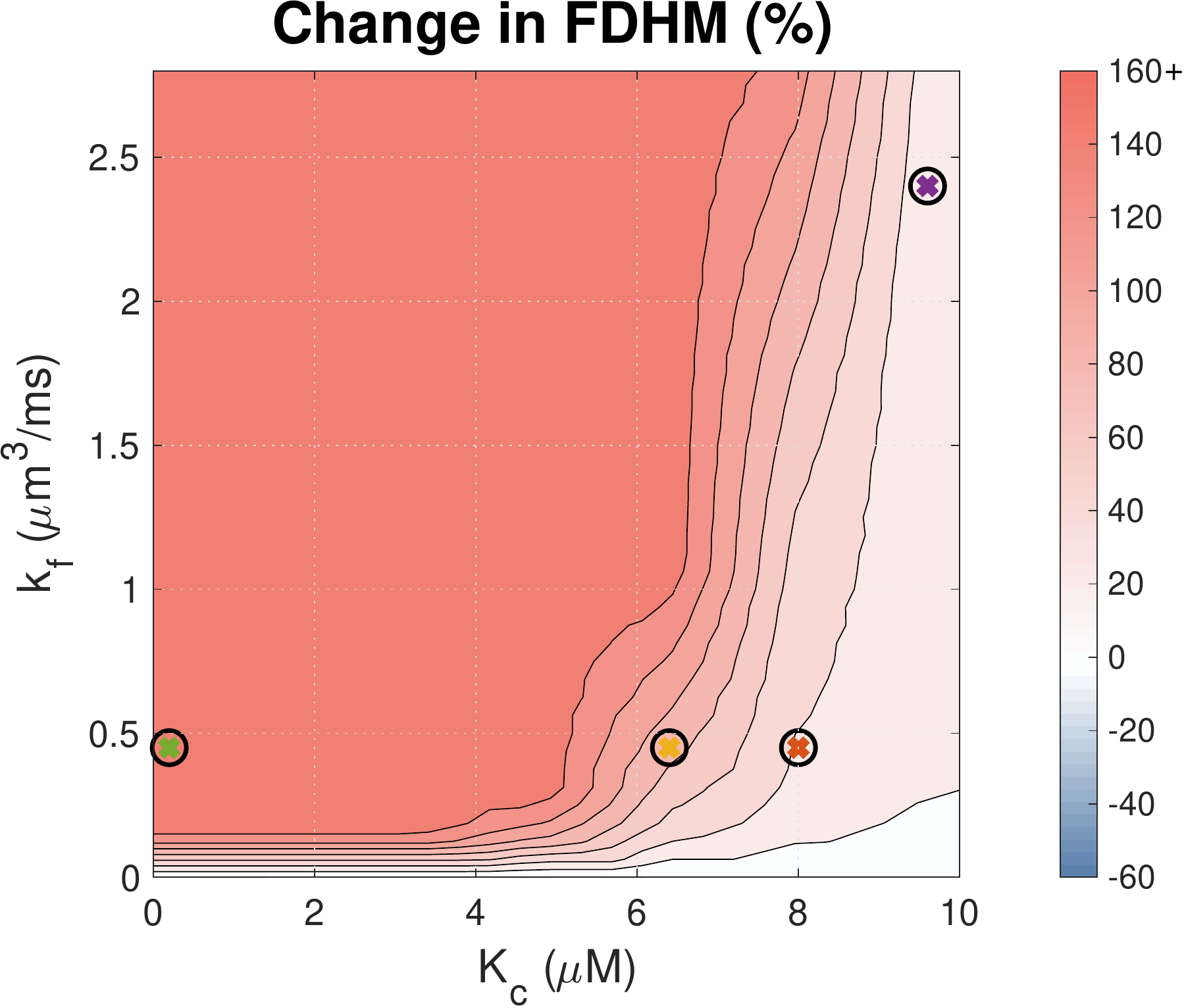}
		\put(-215,200){\Large B} 
	\end{subfigure}
	~\\
	\begin{subfigure}[b]{0.45\textwidth}
		\includegraphics[width=1.\textwidth]{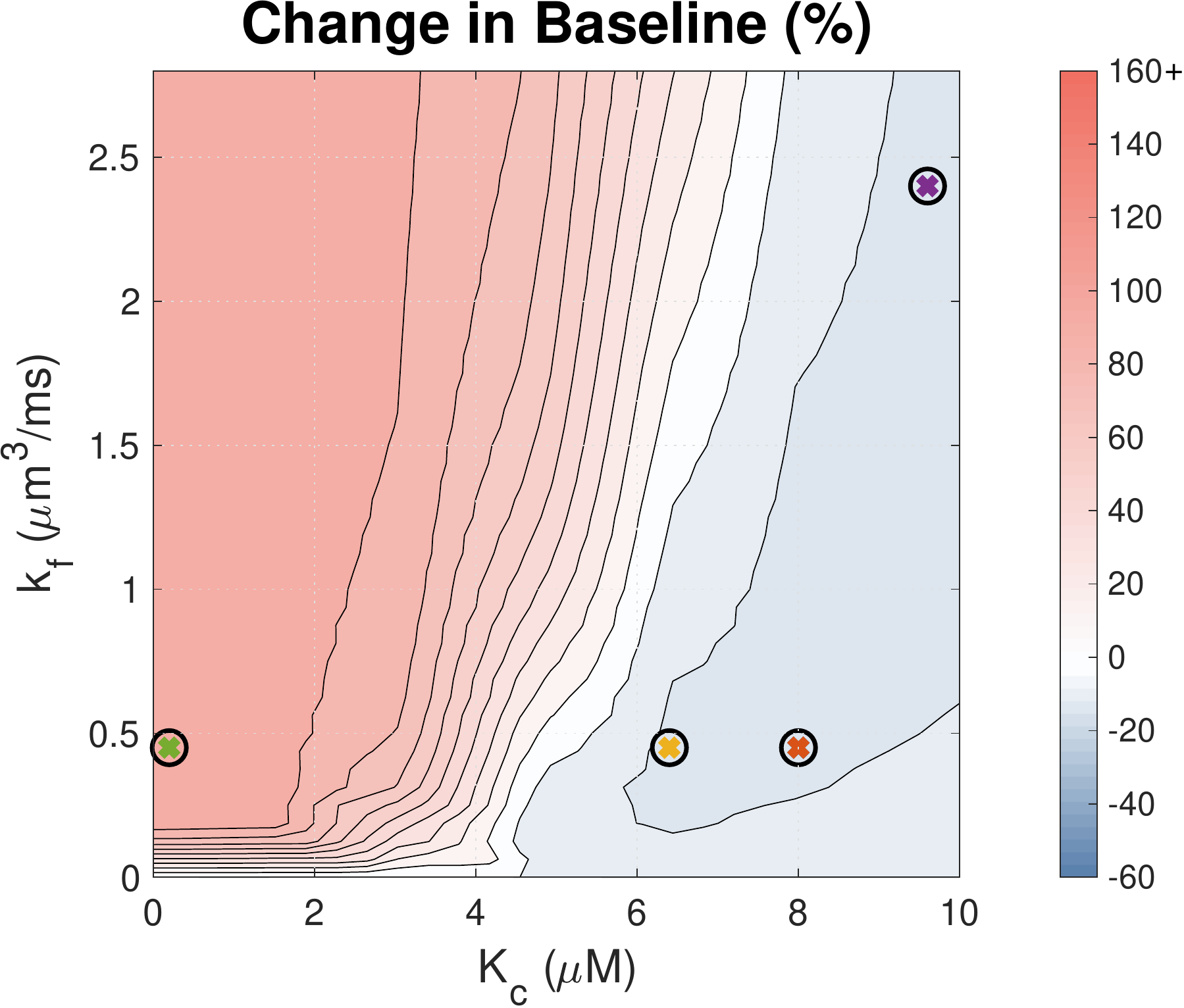}
		\put(-215,200){\Large C}
	\end{subfigure}
	\begin{subfigure}[b]{0.45\textwidth}\hspace{5mm}
		\includegraphics[width=0.84\textwidth]{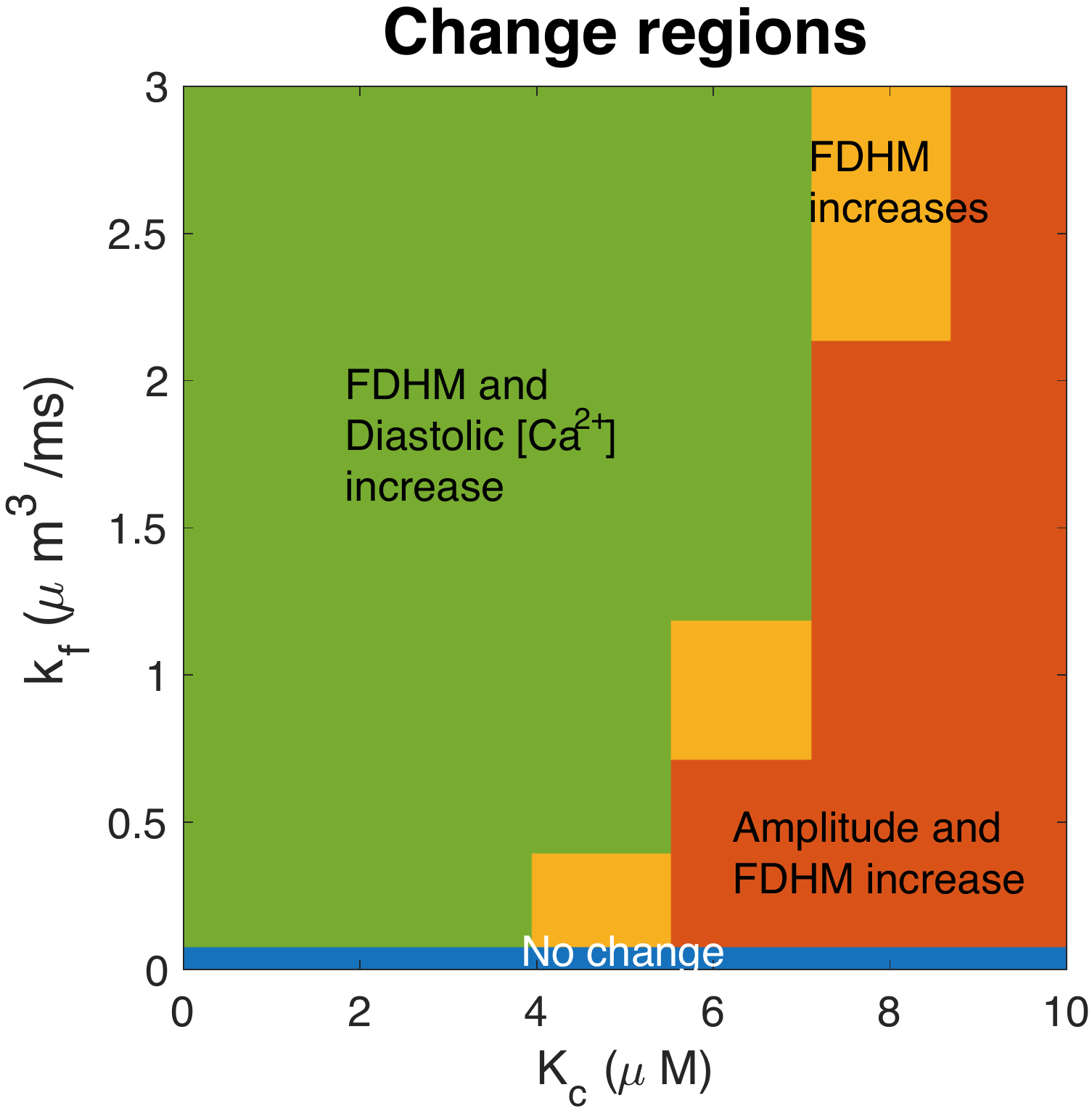}
		\put(-190,200){\Large D}
	\end{subfigure}
		\caption{Effect of maximum \ipr flux $k_{\rm f}$ and the Ca$^{2+}$-sensitivity parameter $K_{\rm c}$ on the Ca$^{2+}$ transient at $1$ Hz. Maximum \ipr flux has the greatest impact on transient duration. In these simulations [\ipns] $=10$ $\mu$M. See Figure \ref{fig:flux_sn} for simulated transients at parameters indicated by crosses.}
	\label{fig:effect_kc_kf}
\end{figure}

\begin{figure}[htp]
	\centering
	\begin{subfigure}[b]{0.45\textwidth}
		\includegraphics[width=1.\textwidth]{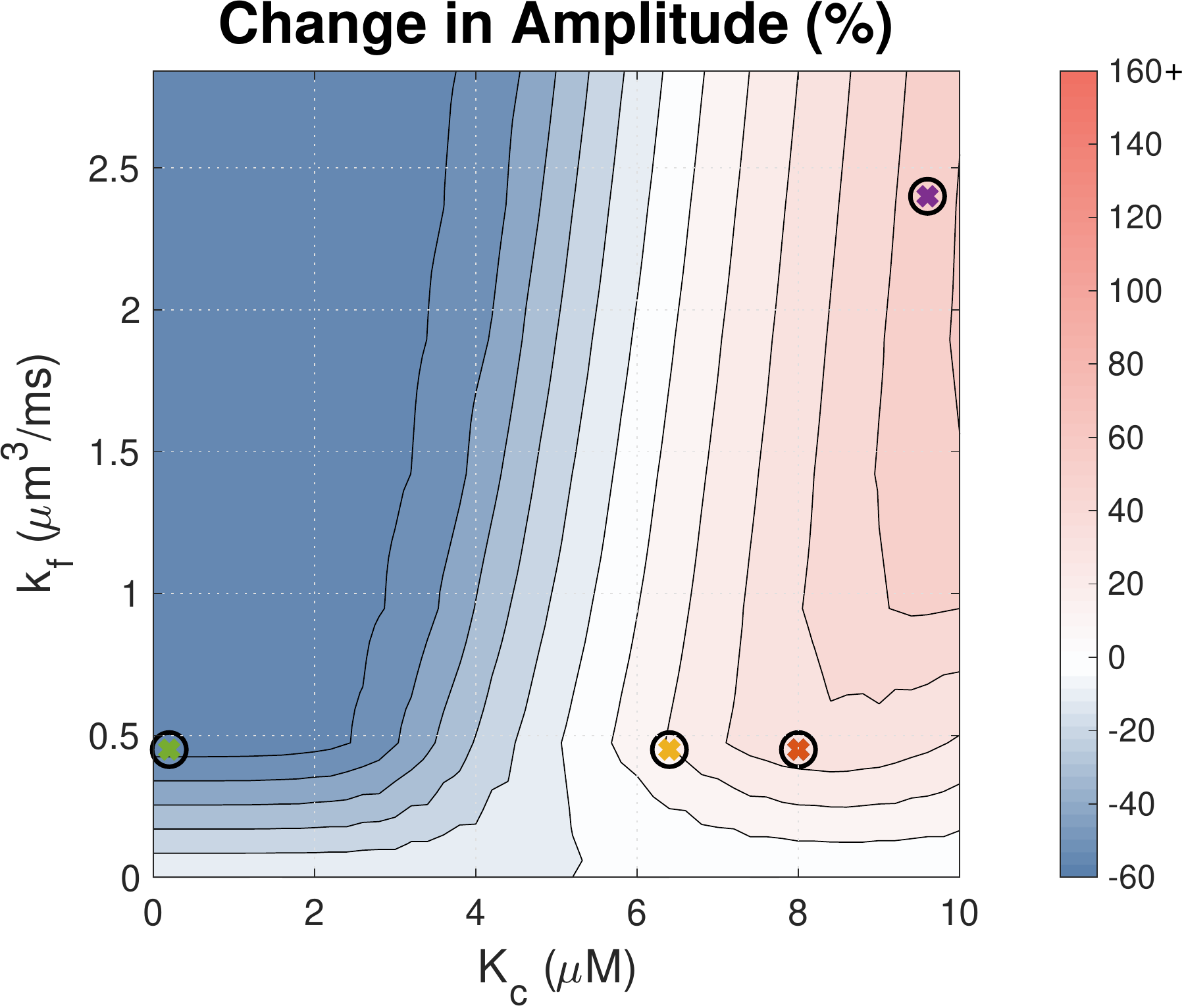}
		\put(-215,200){\Large A}
	\end{subfigure}
	\begin{subfigure}[b]{0.45\textwidth}\hspace{5mm}
		\includegraphics[width=1.\textwidth]{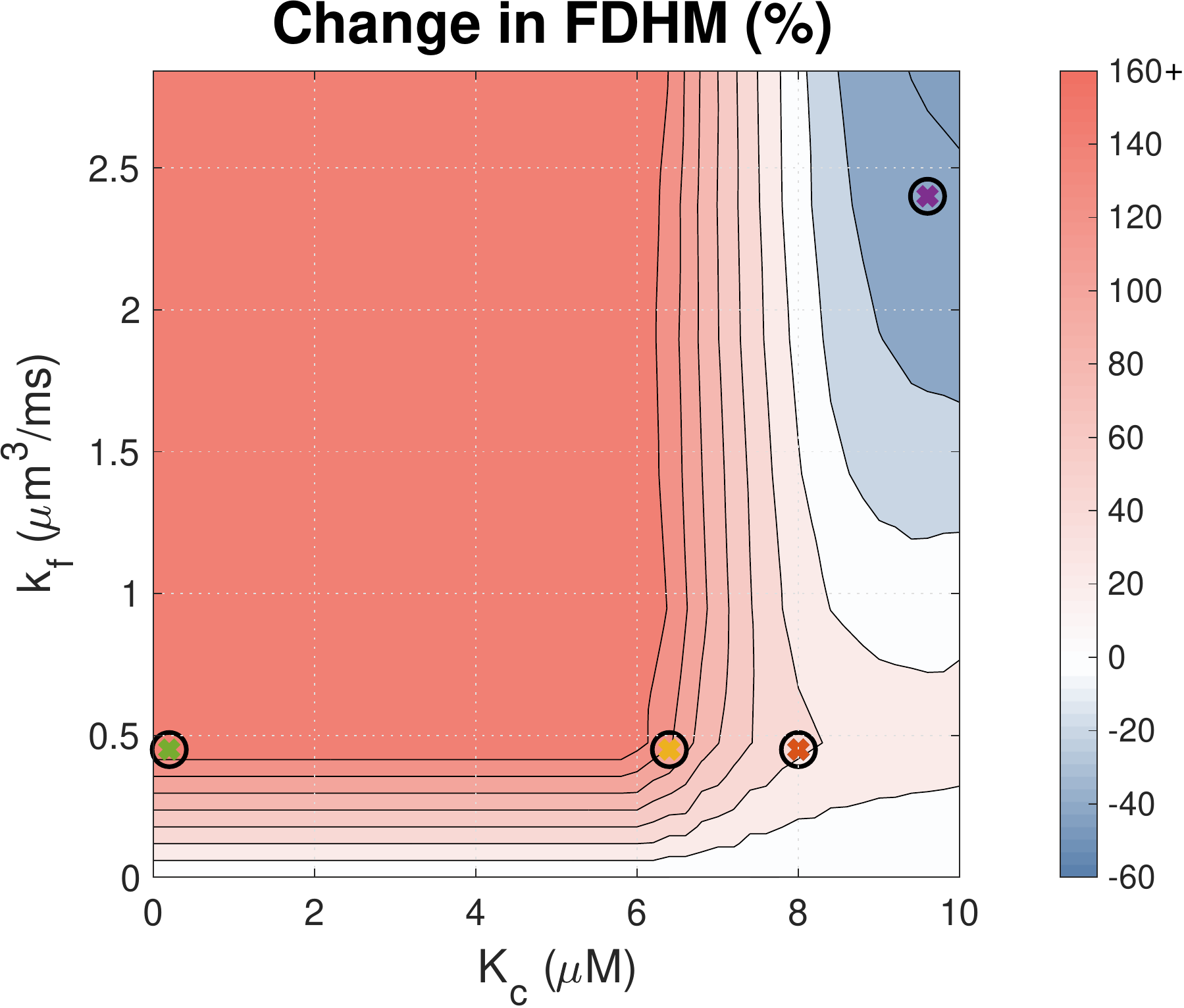}
		\put(-215,200){\Large B}
	\end{subfigure}
	~\\
	\begin{subfigure}[b]{0.45\textwidth}
		\includegraphics[width=1.\textwidth]{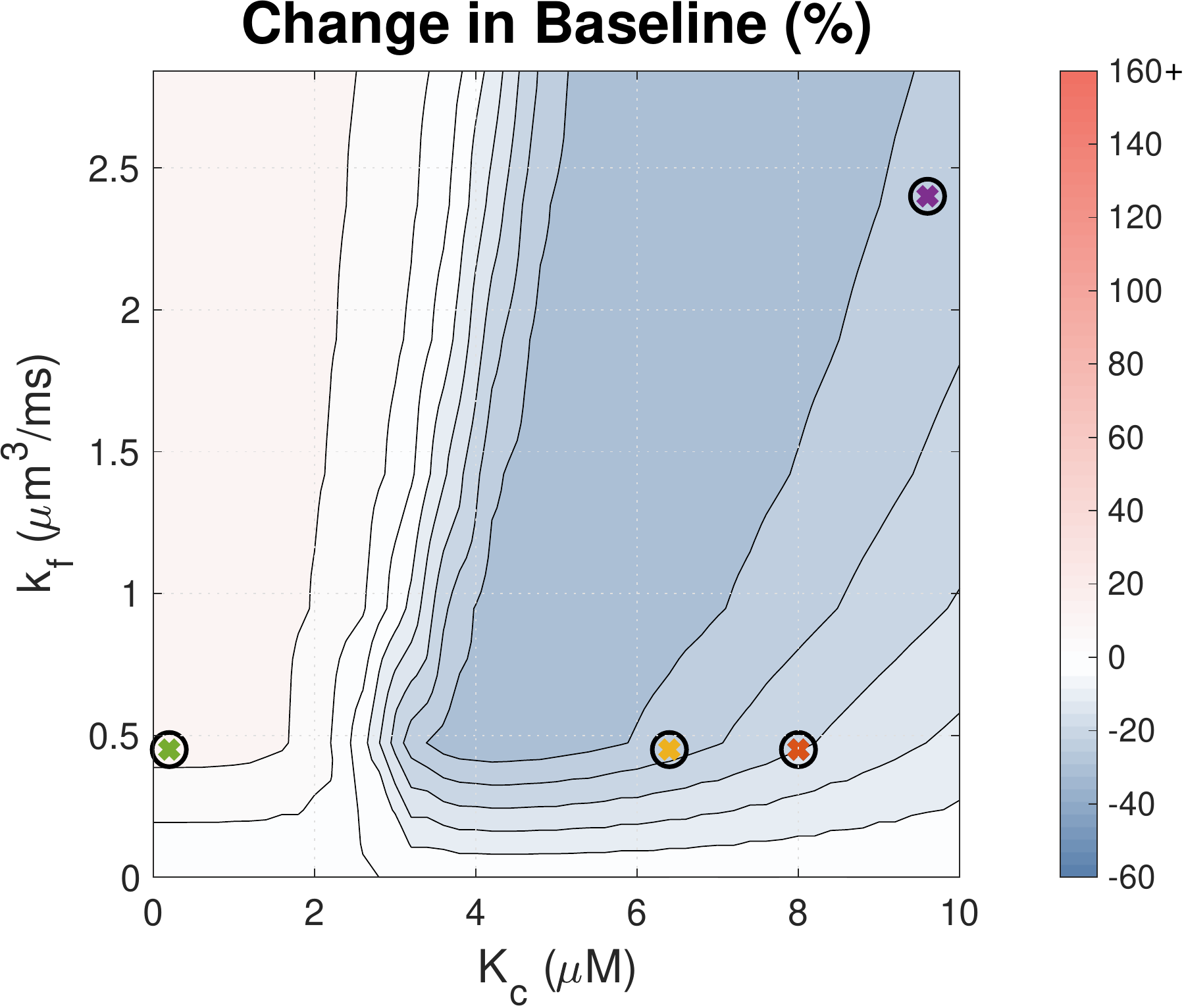}
		\put(-215,200){\Large C}
	\end{subfigure}
	\begin{subfigure}[b]{0.45\textwidth}\hspace{5mm}
		\includegraphics[width=0.84\textwidth]{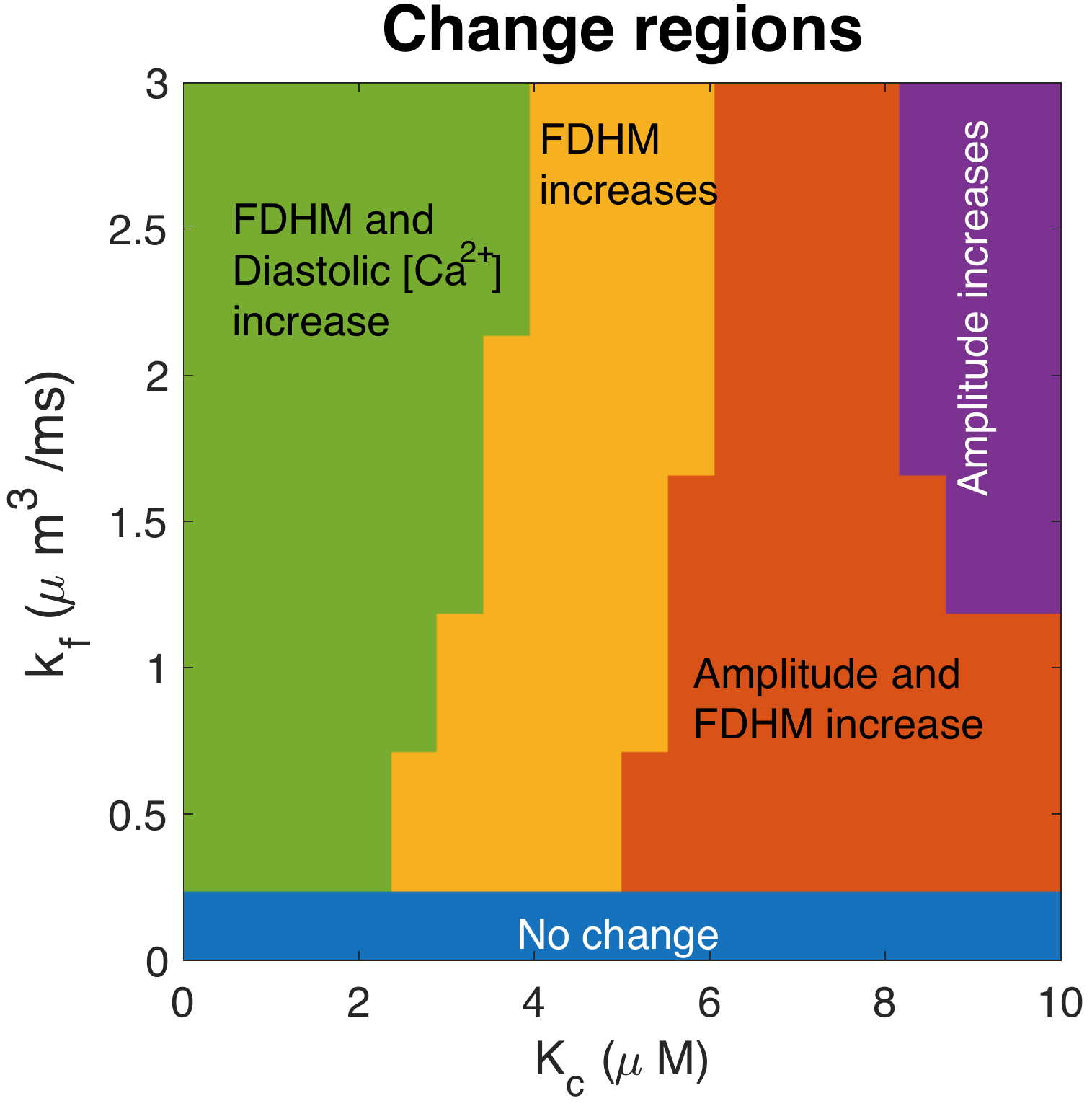}
		\put(-190,200){\Large D}
	\end{subfigure}
		\caption{Effect of maximum \ipr flux $k_{\rm f}$ and the Ca$^{2+}$ sensitivity parameter $K_{\rm c}$ on the Ca$^{2+}$ transient at $0.3$ Hz. Maximum \ipr flux has the greatest impact on transient duration. In these simulations [\ipns] $=10$ $\mu$M. See Figure S3 for simulated transients at parameter values indicated by crosses.}
	\label{fig:effect_kc_kf_3Hz}
\end{figure}

The parameter sensitivity analysis in Table \ref{tab:var_effects} indicates that maximum flux through \iprns s ($k_{\rm f}$) has the greatest effect on Ca$^{2+}$ transient duration. Therefore we next examined how increased $k_{\rm f}$ values in our model affects the Ca$^{2+}$ transient. Figure \ref{fig:effect_kc_kf}A-C show that for $K_{\rm c}<2$ $\mu$M, increasing $k_{\rm f}$ above $0.45$ $\mu$m$^3$m$s^{-1}$ mostly increases transient duration but has only marginal effects on amplitude and baseline. However for large $K_{\rm c}$, the role of $k_{\rm f}$ in modifying transient shape becomes more noticeable. There is a clear region where amplitude increases (red region), however this is more dependent on $K_{\rm c}$ than $k_{\rm f}$. At 1 Hz, there is no value of $k_{\rm f}$ that reduces transient duration. With \ipr activation the transient duration increases and $k_{\rm f}$ merely determines by how much. However it is of note that, as shown in Figure \ref{fig:effect_kc_kf_3Hz}, at a lower frequency of 0.3 Hz, when $k_{\rm f} > 1.2$ $\mu$m$^3$m$s^{-1}$ and $K_{\rm c}>8$ $\mu$M, there is a decrease in duration of the transient. 

To compare simulation results to experimental observations in Table \ref{tab:exp}, we divided the parameter space shown in Figures \ref{fig:effect_kc_kf}A-C into three regions, shown in Figure \ref{fig:effect_kc_kf}D. The regions in this figure are consistent with the regions labelled in Figure \ref{fig:effect_kc_ip3}D. Figure \ref{fig:effect_kc_kf_3Hz}D shows similar regions corresponding to simulations at 0.3 Hz. It can be seen that at 0.3 Hz, $K_{\rm c}>8$ $\mu$M and \mbox{$k_{\rm f}>1.2$ $\mu$ m$^3$m$s^{-1}$} provide transients with increased amplitude and decreased duration, consistent with rat ET-1 experiments summarized in Table \ref{tab:exp}.  However this value of $k_{\rm f}$ results in an unrealistic flux through \ipr channels. Additionally, \emph{in vivo}, the cell would be paced at a faster frequency and this result is unlikely without the cell being able to return to resting Ca$^{2+}$. We have not been able to identify a parameter set that would provide a simultaneous increase in both amplitude and diastolic \ca.

\subsection*{RyR and \ipr interaction increases the intracellular \ca duty cycle}

Having establishing reasonable  parameters ranges for \ipr activation based on the influence on ECC Ca$^{2+}$ transient properties (amplitude, FDHM, and diastolic \cans), we investigated the possibility that cytosolic Ca$^{2+}$ plays a role in hypertrophic remodelling through changing the duty cycle. Given the time scale involved in hypertrophic remodelling, and the signal integration properties of NFAT, the \ipr-modified cytosolic Ca$^{2+}$ transient could cumulatively encode hypertrophic signalling. Using optogenetic encoding of cytosolic \ca  transients in HeLa cells, \citet{hannanta-anan_optogenetic_2016}
demonstrated that the transcriptional activity of NFAT4 can be up-regulated by increasing cytosolic Ca$^{2+}$ duty cycle. This is a plausible mechanism of signal encoding that is likely to be less susceptible to noise than either amplitude or frequency encoding. Therefore, we examined the cytosolic Ca$^{2+}$ duty cycle as a hypertrophic signalling mechanism. 

\begin{figure}[htp]
	\centering
	\begin{subfigure}[b]{0.485\textwidth}
		\includegraphics[width=1.\textwidth]{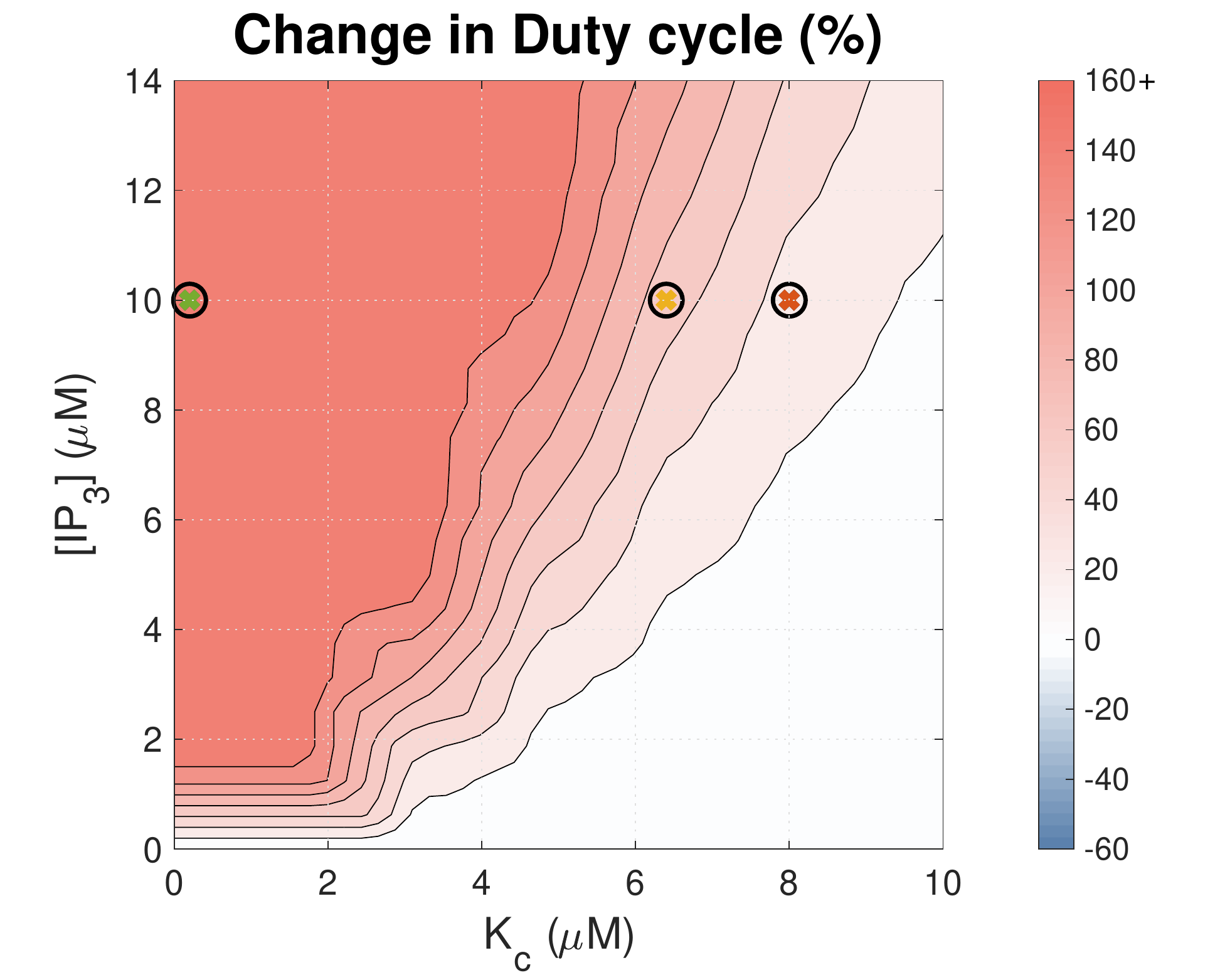}
		\put(-215,200){\Large A}
	\end{subfigure}
	\begin{subfigure}[b]{0.45\textwidth}\hspace{2mm}
		\includegraphics[width=1.\textwidth]{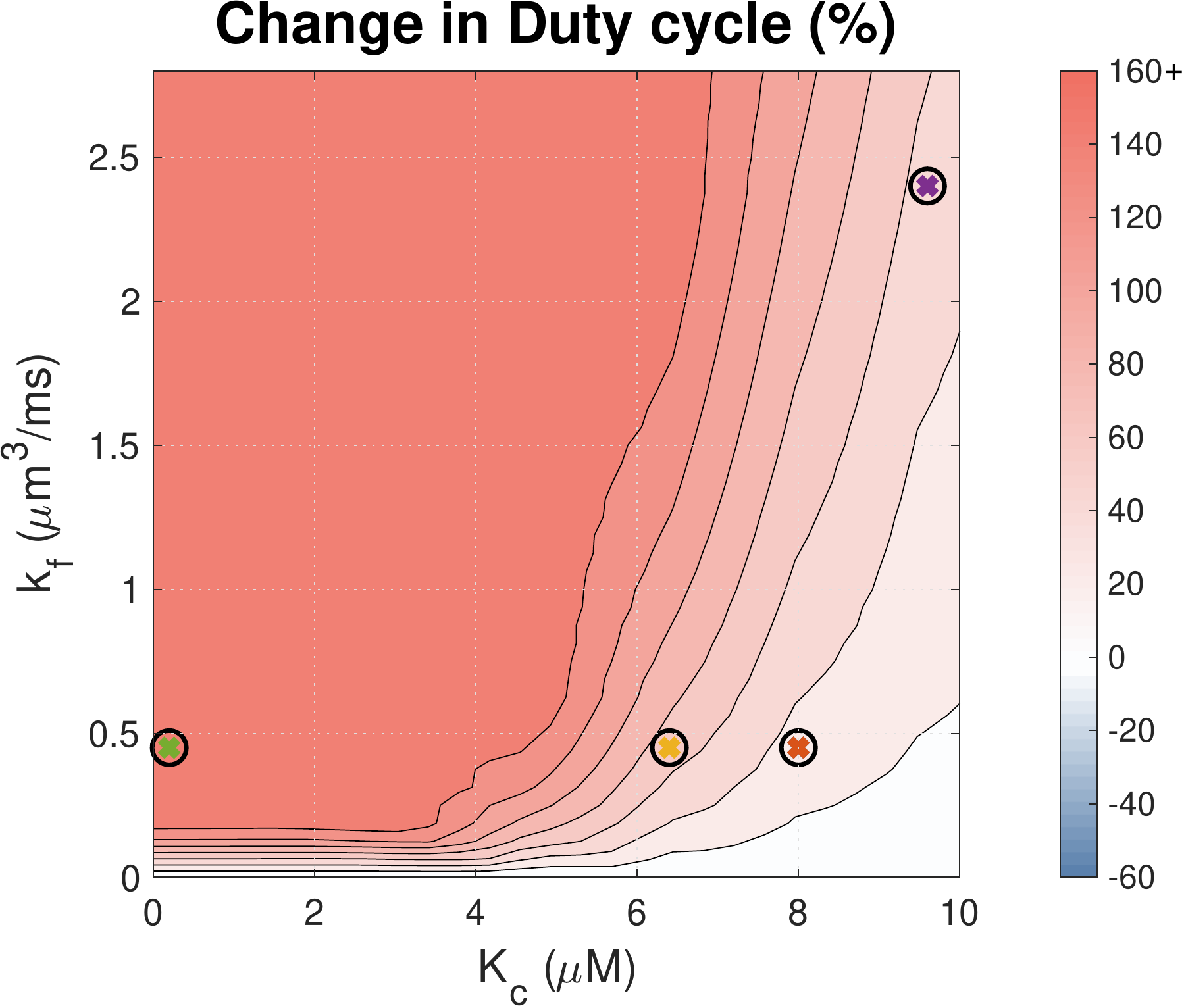}
		\put(-215,200){\Large B}
	\end{subfigure}
	~\\
	\begin{subfigure}[b]{0.45\textwidth}
		\includegraphics[width=1.\textwidth]{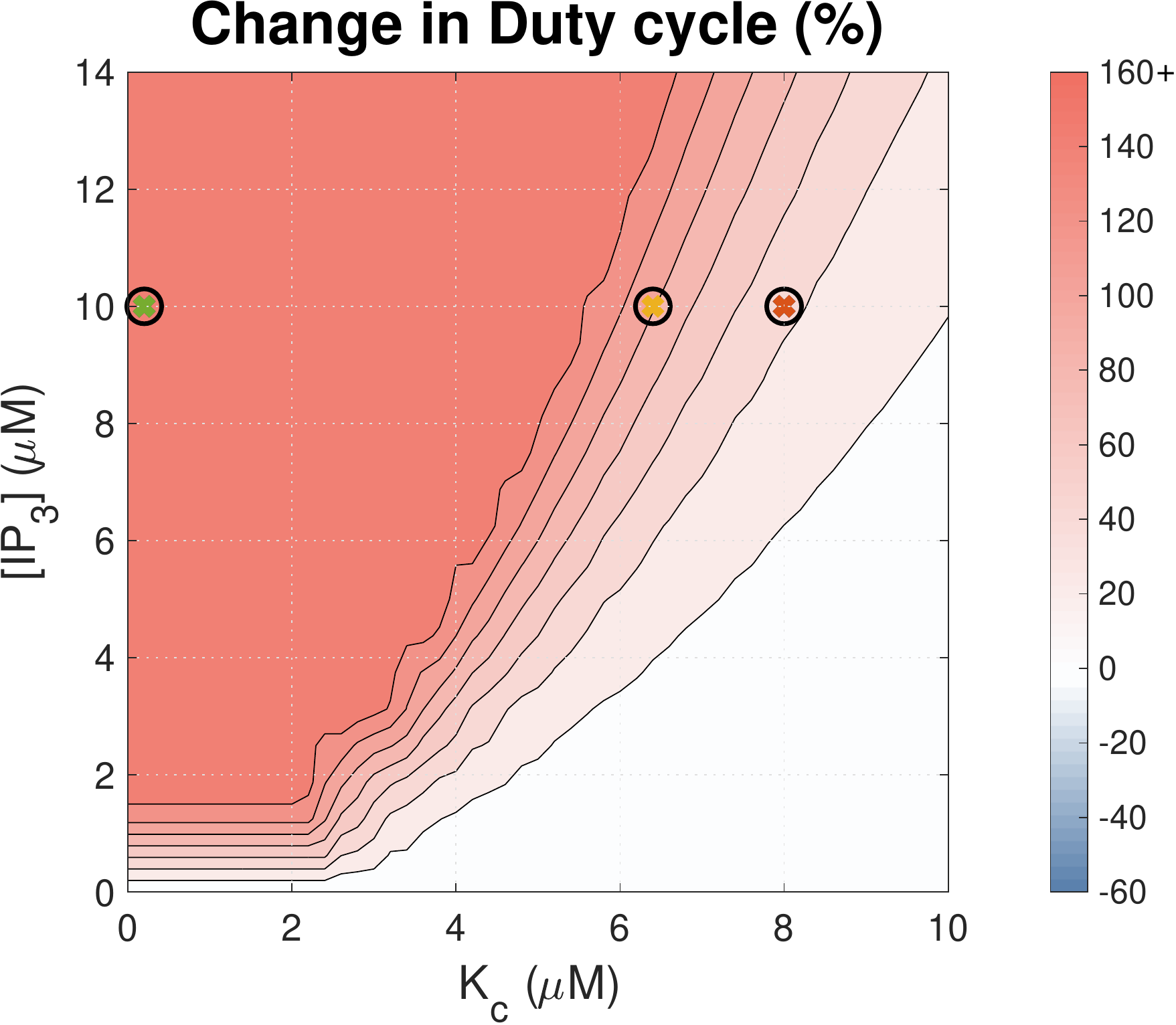}
		\put(-215,200){\Large C}
	\end{subfigure}
	\begin{subfigure}[b]{0.45\textwidth}\hspace{5mm}
		\includegraphics[width=1.\textwidth]{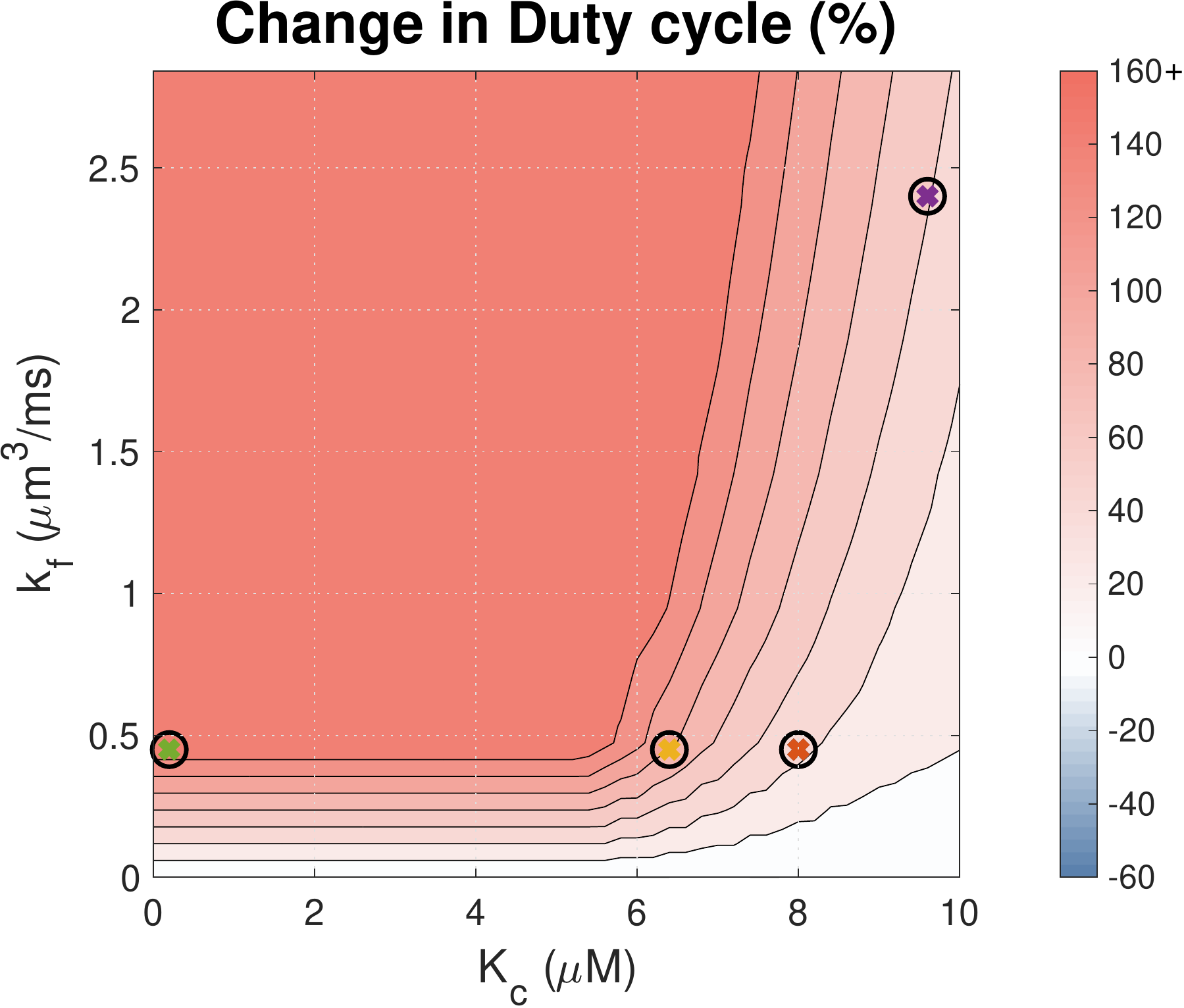}
		\put(-215,200){\Large D}
	\end{subfigure}
    \caption{Effects on the Ca$^{2+}$ transient duty cycle of (A) \ip concentration and the Ca$^{2+}$ sensitivity parameter $K_{\rm c}$ with pacing frequency 1 Hz; (B) of maximum \ipr flux $k_{\rm f}$ and $K_{\rm c}$ with pacing frequency 1 Hz; (C) of \ip concentration and $K_{\rm c}$ with pacing frequency 0.3 Hz; and (D) of maximum \ipr flux $k_{\rm f}$ and the $K_{\rm c}$ at pacing frequency 0.3 Hz. The colour bar indicates the \% change from a simulation run with identical parameters but no \ipr channels. The coloured crosses indicate the parameters used for the corresponding plots in Figure~\ref{fig:flux_sn}. \citet{hannanta-anan_optogenetic_2016} report a transcription rate increase of approximately 30\% with a duty cycle increase of 50\% in Figure 2 of their paper. The duty cycle of the Ca$^{2+}$ transient when \iprs are inactive is 0.127.}
	\label{fig:dc}
\end{figure}

We calculated the duty cycle for the \ca transients in the plausible parameter ranges for \ipr activation as the ratio between the area under the \ca transient curve and the area of the bounded box defined by the amplitude and period of the \ca transient (shown in Figure \ref{fig:cartoontrans}). Figure \ref{fig:dc} shows the effects of [\ipns], $k_{\rm f}$, and $K_{\rm c}$ on the duty cycle of the cytosolic Ca$^{2+}$ transient. The Figure shows that the Ca$^{2+}$ duty cycle increases with \ipr activation across the broad parameter range shown.

%On the other hand, our analysis suggests that high $k_{\rm f}$ values can cause a decrease in the transient duration (see Figure \ref{fig:effect_kc_kf_3Hz}B), but despite this, the duty cycle nevertheless increases in all investigated $K_{\rm c}, k_{\rm f}$ parameter sets. These observations suggest that \ipr can increase duty cycle, and furthermore that for a broad parameter range, containing the plausible physiological range, \ip does cause an increase in the duty cycle.

% 	A \raisebox{-0.8\height}{\includegraphics[scale=0.55,angle=0]{auc-dc}}
% 	B\raisebox{-0.8\height}{\includegraphics[scale=0.6,angle=0]{bar_dc}}
% 	\caption{\textbf{A} Comparison of transients (blue lines), area under the curve, and duty cycle ($\gamma$) in a regime where \iprs have no effect on transient amplitude. Both increased numbers of \iprs and increased \ip concentration increase the duty cycle. \textbf{B} Duty cycle increases with both [\ip] and \ipr numbers.}
\subsection*{\mynotes{NFAT activation increases with an increase in calcium duty cycle}}
\begin{figure}[ht!]
\centering
\includegraphics[width=0.5\textwidth]{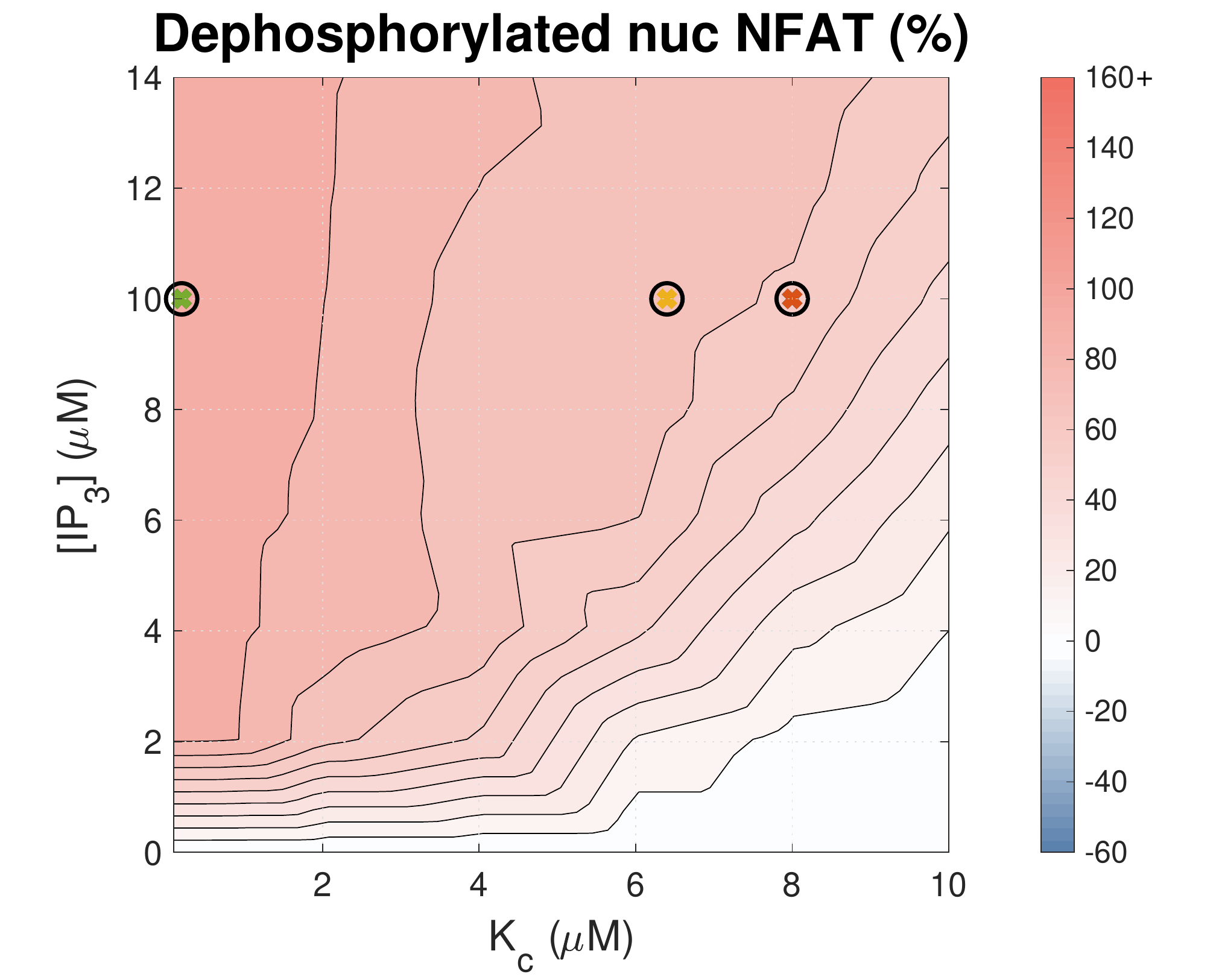} 
\caption{Effect of [\ipns] and $K_{\rm c}$ on the concentration of dephosphorylated nuclear NFAT (NFAT$_{n}$). Simulations were paced at 1 Hz. The colour bar indicates the $\%$ change from a simulation run with identical parameters but no \ipns.}
\label{fig:NFAT}
\end{figure}

\mynotes{Having established that \ipr activation results in increased calcium transient duty cycle, we coupled the model of cytosolic ARVM calcium dynamics to the  model of NFAT activation developed by \citet{cooling_sensitivity_2009}. We then tested the effect of varying \ip concentration over a range of \ipr parameter values on the proportion of dephosphorylated nuclear NFAT compared to that in the phosphorylated inactive state in the cytosol (Figure \ref{fig:NFAT}). These simulation data clearly show that increased \ip and alteration in \ca transient duty cycle positively influences NFAT activation and thus provides a mechanism to couple \ipns-induced \ca release and activation of hypertrophic gene expression.}

\section*{Discussion}
Here we have presented what is, to our knowledge, the first modelling study to investigate the effect of \ipr channel activity on the cardiac ECC Ca$^{2+}$ transient and possible information encoding mechanisms. We extended a well-established model of the ECC \ca transient by \citet{hinch_simplified_2004} to include a model of \ipr activation and \ca release. The model, upon \ipr activation, simulates the influence of \ipr activation on \ca transients in non-hypertrophic adult rat left ventricular cardiac myocytes.

Parameter sensitivity analysis (Table \ref{tab:var_effects}) showed the maximal \ipns-induced \ca release through individual \ipr ($k_{\rm f}$) had the greatest influence on the \ca transient duration and duty cycle. [\ipns] had the biggest influence on the \ca amplitude and diastolic \ca concentration. We found that under fixed maximum \ipr flux, $k_{\rm f}=0.45$ $\mu$m$^3$m$s^{-1}$, \ipr activation increases the duration of the Ca$^{2+}$ transient, but \ca amplitude is \ipns-dependent. The \ca transient duration can be reduced only by increasing $k_{\rm f}$ to physiologically unrealistic values. 

The finding that the \ca transient duty cycle increases with [\ipns] (see Figure~\ref{fig:dc}) provides a \mynotes{plausible} explanation for the mechanism by which \ipns-dependent \ca release from \iprs can enhance pro-hypertrophic NFAT activity.

%The model is able to replicate several experimentally observed effects of varying \ipr activation and \ipns concentration on the ECC Ca$^{2+}$ transient.% 

\subsection*{Does \ipns-induced \ca release modify the ECC transient?}
Figures \ref{fig:effect_kc_ip3}, \ref{fig:effect_kc_kf} and \ref{fig:effect_kc_kf_3Hz} show that \iprs can influence the ECC \ca transient and the effect is dependent on the \ipr properties and \ip concentration. 
%
%The most consistent influence of \ipr activation from our model simulations is the increase in full duration half max of the \ca transient across all parameter ranges that simulate different \ipr gating kinetics.  
%
Our model simulations predict that \ca transient amplitude increases approximately 15\% when \ipr properties are such that \iprs remain inhibited from opening at diastolic \ca but release \ca once RyRs are activated and remain open when \ca concentration is above 1 $\mu$M. The \ipr parameter combination marked by a red cross in the contour plots is a representative example of this type of effect of \iprsns. There is also a narrow parameter range at [\ipns] of 10 $\mu$M ($K_{\rm h}=2.2$ $\mu$M, $K_{\rm c}=$ 6 $\mu$M) where the amplitude does not change more than 5\% (see Figure \ref{fig:effect_kc_ip3}). The orange cross marks an example of \ipr effects in this parameter range. These simulation predictions are consistent with the experimental studies that either show increased amplitude or no change in amplitude (Table \ref{tab:exp}). 
%In rat ventricular myocytes, \citet{higazi_endothelin-1-stimulated_2009} showed that the \ca transient amplitude in the cytosol did not change in both adult and neonatal hearts shortly  ($\sim$50 seconds) after ET-1 stimulation. \citet{harzheim_increased_2009} similarly reported no change in transient amplitude after 15 minutes of 100 nM ET-1 stimulation nor after stimulation with \ip ester at 10 $\mu$M concentration. A close examination of the  data presented in \cite{harzheim_increased_2009} reveals that transient amplitude exhibited subtle increase in amplitude, mean fold-change between 1.0 and 1.2, that was deemed statistically insignificant when compared to amplitudes in cells with no hypertrophic stimulation.

Model simulations predict that \ipr activation only increases diastolic \caconc when \ipr are open at resting \caconc of $\sim$0.1 $\mu$M (see Figure \ref{fig:effect_kc_ip3}D). \citet{harzheim_increased_2009} reported no measurable differences in diastolic \caconc between \mynotes{ARVMs stimulated with an agonist, known to induce hypertrophy in healthy ARVMs, and those treated with a saline buffer (although effects have been observe in disease ventricular cardiomyocytes and atrial cardiomyocytes).} 
%plots showing change in diastolic \caconc in \cite{harzheim_increased_2009} report 1.0 fold-change before and after ET-1 and \ip stimulation. 
Examination of simulated \ca transients within a regime that results in diastolic \caconc  increase (green traces in Figure \ref{fig:flux_sn}) shows that the transients do not resemble any of the observed experimental measurements in the literature. Therefore the comparison of  model simulations and experimental measurements of diastolic \caconc and \ca transient amplitude  suggest that the most likely regime of \ipr activation lies between the orange and red regions in Figure \ref{fig:effect_kc_ip3}D. Using these comparisons we propose that \ipr activation makes modest changes to the ECC \ca transient which are often hidden within the measurement variability in experiments.

\subsection*{The biological significance of the duty cycle}
We showed that while amplitude, duration, and diastolic \ca can increase or decrease depending on \ipr parameter values and pacing frequency, the duty cycle, as defined by \citet{hannanta-anan_optogenetic_2016} always increases with \ipns, consistent with effects seen in \cite{proven_inositol_2006}. The implication of this observation is that \ipr activation is sufficient to provide a signal to drive NFAT nuclear translocation and hence hypertrophic gene expression in the manner described by \citet{hannanta-anan_optogenetic_2016}.

\citet{hannanta-anan_optogenetic_2016} found that, when comparing Ca$^{2+}$ oscillations of the same amplitude, oscillations with greater duty cycle had a greater effect on NFAT dephosphorylation and translocation to the nucleus. In their study, duty cycle, $\gamma$, was calculated as the area under the curve, $U$, divided by the maximum area under the curve (for Ca$^{2+}$ oscillations of the same amplitude, $A$, and period of oscillation, $T$), i.e. $\gamma = U/AT$ (see Figure \ref{fig:cartoontrans}A). An alternative definition is $\gamma = \Delta/T$, where $\Delta$ is  the transient duration and $T$ the period of oscillation. This alternate formulation is used by \citet{tomida_nfat_2003} and \citet{salazar_decoding_2008} but is less well defined for analogue signals. The duty cycle in Figure \ref{fig:dc} was calculated using the former definition. This can be compared with the latter definition when remembering that duty cycle will now vary with FDHM (Figures \ref{fig:effect_kc_ip3}C and \ref{fig:effect_kc_kf}C).

The duty cycle in this system essentially reflects the fraction of each period of the Ca$^{2+}$ cycle for which cytosolic Ca$^{2+}$ is sufficiently elevated to affect the downstream proteins in the CnA/NFAT signalling pathway. The greater sensitivity of NFAT to Ca$^{2+}$  oscillations with sustained elevation in intracellular Ca$^{2+}$ is well established \cite{dolmetsch_differential_1997,colella_ca2+_2008,rinne_isoform-_2010}.  While it is difficult to determine where this threshold is, NFAT is a Ca$^{2+}$ integrator and a clear correlation has been found between Ca$^{2+}$ duty cycle and NFAT activation \cite{hannanta-anan_optogenetic_2016}.  Increasing duty cycle increases the time NFAT spends in the dephosphorylated state, which is required to both enter and maintain it in the nucleus and hence effect transcription \cite{feske2000}; NFAT responds to changes in duty cycle while being insensitive to both amplitude and frequency changes. \mynotes{We see in simulations too that the proportion of NFAT that is in the dephosphorylated nuclear state is highest when the duty cycle of the \ca transient is high (Figures \ref{fig:dc} and \ref{fig:NFAT}).} 

In experiments, \ip stimulation has been shown to lead to an increase in systolic Ca$^{2+}$ in cardiac cells, but significant change in duration has not been reported
% not sure this addition from Llew is right??
(although as in \citet{harzheim_increased_2009} and \citet{proven_inositol_2006}, increased \mynotes{spontaneous calcium transients} are observed which could function to prolong the duration of the Ca$^{2+}$ transient). Based on the definition of the duty cycle \mynotes{presented in \citet{hannanta-anan_optogenetic_2016}, there is a negative effect on duty cycle, and hence NFAT activation, when \ca transient amplitude is increased. However, within the physiologically plausible parameter range we find that simulations with increased \ca transient amplitude also have increased transient duration.} We postulate that NFAT may be responsive to the Ca$^{2+}$ transient through the latter definition of the duty cycle -- i.e. the duration of time that Ca$^{2+}$ is elevated over a threshold divided by the period. This is more consistent with both the biological mechanism and the potential increase in peak Ca$^{2+}$ concentration in the hypertrophic pathway\mynotes{, which may be a side-effect of a corresponding increase in duration over this threshold}. Further research, both theoretical and experimental, is required in order to determine the validity of this assumption.

\mynotes{Figure \ref{fig:NFAT} shows a strong correlation between [\cans]-dependent NFAT activation and \ca transient duty cycle in the \citet{cooling_sensitivity_2009} model: the correspondence between Figure \ref{fig:dc}A and Figure \ref{fig:NFAT} is striking. A caveat, however, is that the original \citet{cooling_sensitivity_2009} study showed that the NFAT model is also sensitive to any average increase in cytosolic calcium. Therefore, while increasing \ca transient duty cycle is shown to be sufficient for NFAT activation in this model, further experimental validation is required to confirm this mechanism in cardiomyocytes.}

\subsection*{Experimental evidence of an \ipns-induced increase in calcium duty cycle?}
An increase in duty cycle without an increase in frequency requires an increase in transient duration. While this increase is observed in our simulations for a broad range of parameter values, it has not however been reported in experiments involving \ip stimulation. The possible reasons for this are many and varied, however, as discussed earlier, using different \ip concentrations to those that occur \emph{in vivo} may result in different effects on the shape of the Ca$^{2+}$ oscillations, leading to inconsistent observations. Furthermore, small variations in \ca concentrations may not be experimentally discernible, or may be hidden by the effect of \cans-sensitive dyes \cite{sparrow_measurement_2019}. A small, but prolonged variation in transient duration can produce a comparatively large change in duty cycle. Hence it remains to be confirmed experimentally whether \iprns-dependent \ca flux does indeed lead to an increased \ca duty cycle in cardiomyocytes.

% \mynotes{
% \subsection*{\ipr flux}
% \ref{fig:flux_sn} shows the fluxes through various calcium channels for four parameter sets.
% }

%\mynotes{Much of the \textit{in vitro }transient data described in the literature was gathered from whole cell experiments in which cardiomyocytes were stimulated with an esterified, cell permeant form of \ip (InsP3-BM) \cite{proven_inositol_2006,harzheim_increased_2009}. The exact \ip concentration that \ipr channels see in these experiments is not clear. It has been found that ET-1 stimulation of cardiomyocytes produces a similar positive inotropic effect on the Ca$^{2+}$ transient to that produced by \ip stimulation, albeit reported more prominently in atrial myocytes than ventricular \cite{lipp2000}. \citet{moravec_endothelin_1989} found that the maximal positive inotropic effect produced by ET-1 on rat cardiomyocytes could be observed at concentrations of 50nM. A report by \citet{remus2006} found that free \ip rises to only 30nM in rat ventricular myocytes after stimulation with 100nM ET-1 -- far from saturation \cite{ramos-franco2000} and far lower than the \ip concentrations used in many of the experiments that report cardiac Ca$^{2+}$ transients. However, it is also possible that these global measurements of \ip may not reflect local concentrations at \ipr channels. The G-coupled receptors that sense ET-1 and release \ip are located in the t-tubules which could result in lower concentrations of ET-1 at the receptors and/or, if located near the dyads, could result in higher \ip at the \ipr than that recorded.}

\subsection*{Limitations of the study}
\mynotes{In this study we have considered generation of voltage-driven cytosolic Ca$^{2+}$ transients using deterministic models of each ion channel in a compartmental model. There are several physiological features of cardiomyocyte \ca dynamics which are not represented, and hence not considered in this approach. In particular, our model does not represent any of the stochastic events associated with \ipr channels. Further modelling of combined stochastic channel gating may be necessary to elucidate the entire impact of \ipr interaction with the cytosolic \ca machinery.} While cell structure is known to play a role in cardiac Ca$^{2+}$ dynamics \cite{gaur_multiscale_2011,rajagopal_examination_2015,ladd_assessing_2019}, effects beyond the synchronising function of the dyad are not considered in this compartmental study. Furthermore we have not considered the spatial \ipr distribution. Our model is developed primarily using parameters fitted by \citet{hinch_simplified_2004} and \citet{sneyd_dynamical_2017}, and makes no distinction between \ipr channels located within or outside the dyad \cite{mohler_ankyrin-b_2003,mohler_ankyrin-b_2005}. These and other structural features of the cell could alter the \ca available to regulate \ipr channels and may be detected in the Ca$^{2+}$ transient. Distinct effects of \ip signalling in the cytosol and the nucleus are also not considered. Cytosolic \ca is thought to promote translocation of NFAT into the nucleus, while nuclear \ca maintains it there \cite{higazi_endothelin-1-stimulated_2009}. We have only investigated the former role for Ca$^{2+}$ signalling within the CnA/NFAT pathway. 

\mynotes{We have explored model behaviour at pacing frequencies of 1 Hz and 0.3 Hz, rather than higher, more physiological frequencies, primarily because the majority of parameters were derived from in vitro experiments conducted at room temperature. Extrapolation of parameters and hence model behaviour to in vivo temperature and correspondingly higher pacing frequency remains challenging. Therefore model predictions must be interpreted cautiously in relation to higher pacing frequencies.}

Additionally, not all components of this signalling pathway have been considered in this study. \cans/calmodulin-dependent kinases II and Class IIa histone deacetylases, for example, are both known \cans-mediated components of the hypertrophic pathway that are activated by \ip signalling \cite{wu2006} but are not included. Here we have focused only on the impact of \ipr activation on the cytosolic Ca$^{2+}$ dynamics and how this relates to the mechanism of NFAT activation. In order to explore broader context for \ip mediated hypertrophic signalling, it remains to couple this model to upstream events including models of \ip production through activation of cell membrane receptors \cite{cooling_modeling_2007,cooling_modelling_2008}. \mynotes{This would allow the profile and extent of the rise in \ip concentration due to the activation of the hypertrophic pathway in cardiomyocytes to be determined. We have focused on the effect of an elevated \ip concentration of 10 $\mu$M as many experimental studies into the effect of \ip on \ca dynamics use saturating [\ipns]. However, \citet{remus2006} found stimulation of adult cat ventricular myocytes with 100 nM ET-1 induced a cell-averaged increase in \ip concentration of only 10 nM indicating a much lower concentration than used in experiments. This, together with known differences between species, suggests the \ip concentration detected by \ipr receptors in ARVMs in vivo could be lower than the simulated 10 $\mu M$. However we note qualitatively similar effects on the \ca transient in parameter regimes with lower [\ipns] in our model (Figures \ref{fig:effect_kc_ip3} and S2) albeit with more modest effects on the transient shape. Additionally, ET-1 receptors are localised to t-tubule membranes \cite{boivin2003} so \ip may be generated very close to \ipr channels \cite{mohler_ankyrin-b_2005,escobar2011}, increasing the concentration they detect.}

\mynotes{Finally, \iprns-induced \ca release is a part of a larger hypertrophic signalling network. It remains to couple this model to other signalling pathways involved in bringing about hypertrophic remodelling \cite{ryall_network_2012}. How cytosolic \ca interacts with nuclear \ca in regulation of NFAT nuclear residence and activity also remains to be determined. }

\subsection*{Conclusion}
The sensitivity of NFAT translocation to the Ca$^{2+}$ duty cycle demonstrated by \citet{hannanta-anan_optogenetic_2016} raises the question as to whether \ipr flux can increase the Ca$^{2+}$ duty cycle in cardiomyocytes during hypertrophic signalling. Here we have shown using mathematical modelling that an increase in cytosolic Ca$^{2+}$ transient duration can occur following addition of \ipns, and furthermore that this increase is sufficient to increase NFAT activation. Together, these results suggest a plausible mechanism for hypertrophic signalling via \ipr activation in cardiomyocytes. While it cannot be ruled out that a significant role is played by components of this pathway that are not considered here, the computational evidence provided in this study, along with the previous experimental findings, suggests encoding of the hypertrophic signal through alteration of the duration of cytosolic Ca$^{2+}$ oscillations to be a feasible mechanism for \ipns-dependent hypertrophic signalling.

% \section*{Supporting Material}

\section*{Author Contributions}
EJC, VR, HLR, CS, and GB conceived of the study; EJC and VR supervised the project; HH, AT, VR and EJC developed the modelling approach. HH implemented the simulations. HLR, CS, and GB provided critical feedback. All authors contributed to writing the manuscript.  

\section*{Acknowledgements}
This research was supported in part by the Australian Government through the Australian Research Council Discovery Projects funding scheme (project DP170101358). HLR wishes to acknowledge financial support from the Research Foundation Flanders (FWO) through Project Grant G08861N and Odysseus programme Grant 90663. 
\section*{Supporting Citations}
References \cite{yu_physiome_2011,thomas_comparison_2000} appear in the Supporting Material.

\bibliography{Hunt2020biblio.bib}

\end{document}

\maketitle

\section*{Model Equations}
Model ODEs are described in the main text and as follows:
\begin{align}
    \deriv{[{\rm Ca}^{2+}]_{\rm SR}}{t} &= \frac{V_{\rm myo} }{V_{\rm SR}} \cdot \left(-I_{\rm RyR} + I_{\rm SERCA} - I_{\rm SRl} - I_{\rm IP_3R}\right)\\
    \deriv{\rm TnC}{t} &= I_{\rm TnC}
\end{align}

\subsection*{CaRU model}
We use the reduced, \citet{hinch_simplified_2004}, model of the CaRU as described in \citet{yu_physiome_2011}. The CaRU is modelled as having four states, $z_1,z_2,z_3,z_4$, each describing a different combination of an LTCC and an RyR channel being either open or closed. $J_{Li}$ and $J_{Ri}$ are the total flux through the LTCCs and RyRs in state $i$ respectively.
These equations are detailed below.
\begin{align}
	I_{\rm CaL}&=\frac{N}{V_{\rm myo}}\cdot \left(z_1\cdot J_{L1}+z_2\cdot J_{L2}\right)\\
	I_{\rm RyR}&=\frac{N}{V_{\rm myo}}\cdot\left(z_1\cdot J_{R1}+z_3\cdot J_{R3}\right)
\end{align}

\begin{align}
	\deriv{z_1}{t}&=-(r_1+r_5)\cdot z_1+r_2\cdot z_2+r_6\cdot z_3\\
	\deriv{z_2}{t}&=r_1\cdot z_1-(r_2+r_7)\cdot z_2+r_8\cdot z_4\\
	\deriv{z_3}{t}&=r_5\cdot z_1-(r_6+r_3)\cdot z_3+r_4\cdot z_4\\
	z_4&=1-z_1-z_2-z_3
\end{align}

\begin{align}
	J_{L1}&=J_{Loo}\cdot y_{oo}+J_{Loc}\cdot y_{oc}\\
	J_{L2}&=\frac{J_{Loc}\cdot \alpha_p}{\alpha_p+\alpha_m}\\
	J_{R1}&=y_{oo}\cdot J_{Roo}+J_{Rco}\cdot y_{co}\\
	J_{R3}&=\frac{J_{Rco}\cdot \beta_{pcc}}{\beta_m+\beta_{pcc}}
\end{align}
where the CaRU fluxes are described as
\begin{align}
	J_{Rco} &= \cfrac{J_R\cdot \left([{\rm Ca}^{2+}]_{\rm SR}-[{\rm Ca}^{2+}]_i\right)}{g_D+J_R}\\
  J_{Roo} &= 
    \begin{cases}
        \cfrac{J_R\cdot \left([{\rm Ca}^{2+}]_{\rm SR}-[{\rm Ca}^{2+}]_i+\cfrac{\frac{J_L}{g_D}\cdot {\rm FVRT}_{\rm Ca}}{1-e^{-{\rm FVRT}_{\rm Ca}}}\cdot \left([{\rm Ca}^{2+}]_{\rm SR}-[{\rm Ca}^{2+}]_o\cdot e^{-{\rm FVRT}_{\rm Ca}}\right)\right)}{1+\cfrac{J_R}{g_D}+\cfrac{\frac{J_L}{g_D}\cdot {\rm FVRT}_{\rm Ca}}{1-e^{-{\rm FVRT}_{\rm Ca}}}} & \quad \abs{{\rm FVRT}_{\rm Ca}} > 10^{-5}\\
				&\\
        \cfrac{J_R\cdot \left([{\rm Ca}^{2+}]_{\rm SR}-[{\rm Ca}^{2+}]_i+\cfrac{\frac{J_L}{g_D}\cdot 10^{-5}}{1-e^{-10^{-5}}}\cdot \left([{\rm Ca}^{2+}]_{\rm SR}-[{\rm Ca}^{2+}]_o\cdot e^{-10^{-5}}\right)\right)}{1+\cfrac{J_R}{g_D}+\cfrac{\frac{J_L}{g_D}\cdot 10^{-5}}{1-e^{-10^{-5}}}} &\quad \text{otherwise}
					\end{cases} \\
  J_{Loc} &= \begin{cases}
        \cfrac{\cfrac{J_L\cdot {\rm FVRT}_{\rm Ca}}{1-e^{-{\rm FVRT}_{\rm Ca}}}\cdot \left([{\rm Ca}^{2+}]_o\cdot e^{-{\rm FVRT}_{\rm Ca}}-[{\rm Ca}^{2+}]_i\right)}{1+\cfrac{J_L}{g_D}\cdot \cfrac{{\rm FVRT}_{\rm Ca}}{1-e^{-{\rm FVRT}_{\rm Ca}}}} &\qquad\qquad\qquad\qquad\qquad\qquad\qquad \abs{{\rm FVRT}_{\rm Ca}} > 10^{-5}\\
        \cfrac{\cfrac{J_L\cdot 10^{-5}}{1-e^{-10^{-5}}}\cdot \left([{\rm Ca}^{2+}]_o\cdot e^{-10^{-5}}-[{\rm Ca}^{2+}]_i\right)}{1+\cfrac{J_L}{g_D}\cdot \cfrac{10^{-5}}{1-e^{-10^{-5}}}} &\qquad\qquad\qquad\qquad\qquad\qquad\qquad\text{otherwise}
				\end{cases}\\
  J_{Loo} &= \begin{cases}
          \cfrac{\cfrac{J_L\cdot {\rm FVRT}_{\rm Ca}}{1-e^{-{\rm FVRT}_{\rm Ca}}}\cdot \left([{\rm Ca}^{2+}]_o\cdot e^{-{\rm FVRT}_{\rm Ca}}-[{\rm Ca}^{2+}]_i+\frac{J_R}{g_D}\cdot \left([{\rm Ca}^{2+}]_o\cdot e^{-{\rm FVRT}_{\rm Ca}}-[{\rm Ca}^{2+}]_{\rm SR}\right)\right)}{1+\cfrac{J_R}{g_D}+\cfrac{J_L}{g_D}\cdot \cfrac{{\rm FVRT}_{\rm Ca}}{1-e^{{\rm FVRT}_{\rm Ca}}}}&\abs{{\rm FVRT}_{\rm Ca}} > 10^{-5}\\
          \cfrac{\cfrac{J_L\cdot 10^{-5}}{1-e^{-10^{-5}}}\cdot \left([{\rm Ca}^{2+}]_o\cdot e^{-10^{-5}}-[{\rm Ca}^{2+}]_i+\frac{J_R}{g_D}\cdot \left([{\rm Ca}^{2+}]_o\cdot e^{-10^{-5}}-[{\rm Ca}^{2+}]_{\rm SR}\right)\right)}{1+\cfrac{J_R}{g_D}+\cfrac{J_L}{g_D}\cdot \cfrac{10^{-5}}{1-e^{-10^{-5}}}}&\text{otherwise}
					\end{cases}
\end{align}
where
\begin{align}
	{\rm FVRT}&=\frac{FV}{RT}\\
	{\rm FVRT}_{\rm Ca}&=2{\rm FVRT}
\end{align}
$F$ being the Faraday constant, $V$ the voltage across the cell membrane, described later, $R$ the gas constant and T the temperature. \\

CaRU reduced states:
\begin{align}
    r_1 &= y_{oc}\cdot \mu_{poc}+y_{cc}\cdot \mu_{pcc}\\
    r_2 &= \frac{\alpha_p\cdot \mu_{moc}+\alpha_m\cdot \mu_{mcc}}{\alpha_p+\alpha_m}\\
  r_3 &= \frac{\beta_m\cdot \mu_{pcc}}{\beta_m+\beta_{pcc}}\\
  r_4 &= \mu_{mcc}\\
  r_5 &= y_{co}\cdot \epsilon_{pco}+y_{cc}\cdot \epsilon_{pcc}\\
  r_6 &= \epsilon_m\\
  r_7 &= \frac{\alpha_m\cdot \epsilon_{pcc}}{\alpha_p+\alpha_m}\\
  r_8 &= \epsilon_m
\end{align}
and
\begin{align}
	{\rm expVL} &= e^{\frac{V-V_L}{del_{VL}}}\\
  t_R &= 1.17\cdot t_L\\
  \alpha_p &= \frac{{\rm expVL}}{t_L\cdot ({\rm expVL}+1)}\\
  \alpha_m &= \phi_L/t_L\\
  \beta_{poc} &= \frac{C_{oc}^2}{t_R\cdot (C_{oc}^2+K_{\rm RyR}^2)}\\
  \beta_{pcc} &= \frac{[{\rm Ca}^{2+}]_i}{t_R\cdot ([{\rm Ca}^{2+}]_i^2+K_{\rm RyR}^2)}\\
  \beta_m &= \frac{\phi_R}{t_R}
\end{align}
\begin{align}
  \epsilon_{pco} &= \frac{C_{co}\cdot ({\rm expVL}+a)}{\tau_L\cdot K_L\cdot ({\rm expVL}+1)}\\
  \epsilon_{pcc} &= \frac{[{\rm Ca}^{2+}]_i\cdot ({\rm expVL}+a)}{\tau_L\cdot K_L\cdot ({\rm expVL}+1)}\\
  \epsilon_m &= \frac{b\cdot ({\rm expVL}+a)}{\tau_L\cdot (b\cdot {\rm expVL}+a)}
\end{align}
\begin{align}
  \mu_{poc} &= \frac{(C_{oc}^2+c\cdot K_{\rm RyR}^2)}{\tau_R\cdot (C_{oc}^2+K_{\rm RyR^2)}}\\
  \mu_{pcc} &= \frac{[{\rm Ca}^{2+}]_i^2+c\cdot K_{\rm RyR}^2}{\tau_R\cdot ([{\rm Ca}^{2+}]_i^2+K_{\rm RyR}^2)}\\
  \mu_{moc} &= \frac{\theta_R\cdot d\cdot (C_{oc}^2+c\cdot K_{\rm RyR}^2)}{\tau_R\cdot (d\cdot C_{oc}^2+c\cdot K_{\rm RyR}^2)}\\
  \mu_{mcc} &= \frac{\theta_R\cdot d\cdot ([{\rm Ca}^{2+}]_i^2+c\cdot K_{\rm RyR}^2)}{\tau_R\cdot (d\cdot [{\rm Ca}^{2+}]_i^2+c\cdot K_{\rm RyR}^2)}
\end{align}
CaRU states
\begin{align}
	C_{cc} &= [{\rm Ca}^{2+}]_i\\
  C_{co} &= \frac{[{\rm Ca}^{2+}]_i\cdot g_D+J_R\cdot [{\rm Ca}^{2+}]_{\rm SR}}{g_D+J_R}\\
	C_{oc} &= \begin{cases}
	\cfrac{g_D\cdot [{\rm Ca}^{2+}]_i+\cfrac{J_L\cdot [{\rm Ca}^{2+}]_o\cdot {\rm FVRT}_{\rm Ca}\cdot e^{-{\rm FVRT}_{\rm Ca}}}{1-e^{-{\rm FVRT}_{\rm Ca}}}}{g_D+\cfrac{J_L\cdot {\rm FVRT}_{\rm Ca}}{1-e^{-{\rm FVRT}_{\rm Ca}}}} &\quad \abs{{\rm FVRT}_{\rm Ca}} > 10^{-9}\\
	&\\
	\cfrac{g_D\cdot [{\rm Ca}^{2+}]_i+J_L\cdot [{\rm Ca}^{2+}]_o}{g_D+J_L} &\quad \text{otherwise}
	\end{cases} \\
	  C_{oo} &= \begin{cases}
	\cfrac{g_D\cdot [{\rm Ca}^{2+}]_i+J_R\cdot [{\rm Ca}^{2+}]_{\rm SR}+\cfrac{J_L\cdot [{\rm Ca}^{2+}]_o\cdot {\rm FVRT}_{\rm Ca}\cdot e^{-{\rm FVRT}_{\rm Ca}}}{1-e^{-{\rm FVRT}_{\rm Ca}}}}{g_D+J_R+\cfrac{J_L\cdot {\rm FVRT}_{\rm Ca}}{1-e^{-{\rm FVRT}_{\rm Ca}}}} &\quad \abs{{\rm FVRT}_{\rm Ca}} > 10^{-9}\\
	&\\
	\cfrac{g_D\cdot [{\rm Ca}^{2+}]_i+J_R\cdot [{\rm Ca}^{2+}]_{\rm SR}+J_L\cdot [{\rm Ca}^{2+}]_o}{g_D+J_R+J_L} &\quad \text{otherwise}
	\end{cases}
\end{align}

\begin{align}
	{denom} &= (\alpha_p+\alpha_m)\cdot ((\alpha_m+\beta_m+\beta_{poc})\cdot (\beta_m+\beta_{pcc})+\alpha_p\cdot (\beta_m+\beta_{poc}))
	\end{align}
	\begin{align}
  y_{oc} &= \frac{\alpha_p\cdot \beta_m\cdot (\alpha_p+\alpha_m+\beta_m+\beta_{pcc}}{{denom}}\\
  y_{co} &= \frac{\alpha_m\cdot (\beta_{pcc}\cdot (\alpha_m+\beta_m+\beta_{poc})+\beta_{poc}\cdot \alpha_p)}{{denom}}\\
  y_{oo} &= \frac{\alpha_p\cdot (\beta_{poc}\cdot (\alpha_p+\beta_m+\beta_{pcc})+\beta_{pcc}\cdot \alpha_m)}{{denom}}\\
  y_{cc} &= \frac{\alpha_m\cdot \beta_m\cdot (\alpha_m+\alpha_p+\beta_m+\beta_{poc})}{{denom}}
	\end{align}
% 	\begin{align}
%   y_{ci} &= \frac{\alpha_m}{\alpha_p+\alpha_m)}\\
%   y_{oi} &= \frac{\alpha_p}{\alpha_p+\alpha_m)}\\
%   y_{ic} &= \frac{\beta_m}{\beta_{pcc}+\beta_m)}\\
%   y_{io} &= \frac{\beta_{pcc}}{\beta_{pcc}+\beta_m)}\\
%   y_{ii} &= 1-y_{oc}-y_{co}-y_{oo}-y_{cc}-y_{ci}-y_{ic}-y_{oi}-y_{io}
% \end{align}

% \twocolumn
\subsection*{Extracellular exchange and the cell membrane}
\begin{align}
I_{\rm NCX}&=\frac{g_{\rm NCX}\cdot\left(e^{\eta\cdot {\rm FVRT}}\cdot [\rm Na^+]_i^3\cdot [{\rm Ca}^{2+}]_e-e^{(\eta-1)\cdot {\rm FVRT}}\cdot [\rm Na^+]_e^3\cdot [{\rm Ca}^{2+}]_i\right)}{\left([\rm Na^+]_e^3+K_{\rm mNa}^3\right)\cdot\left([{\rm Ca}^{2+}]_e+K_{\rm mCa}\right)\cdot\left(1+k_{sat}\cdot e^{(\eta-1)\cdot {\rm FVRT}}\right)}\\
I_{\rm PMCA}&=\frac{g_{\rm PMCA}\cdot [{\rm Ca}^{2+}]_{i}}{K_{\rm PMCA}+[{\rm Ca}^{2+}]_{i}}\\
I_{\rm CaB}&=g_{\rm CaB}\cdot\left(E_{\rm Ca}-V\right)
\end{align}
where 
\begin{align}	
	E_{\rm Ca} &= \frac{RT}{2F}\cdot \ln \left( \frac{[{\rm Ca}^{2+}]_o}{[{\rm Ca}^{2+}]_i} \right)
\end{align}

\begin{align}
V&=\begin{cases}
-0.4 \; \text{mod}(t,T_{osc}) &\text{ if} \; \text{mod}(t,T_{osc}) \leq \text{200 ms}\\
\quad V_0 &\text{ otherwise}
\end{cases}
\end{align}
where t is the time since the start of the simulation and $T$\textsubscript{osc} is the period of the driving voltage. The shape of the driving voltage was altered from that described in the original Hinch et al model. Examination of the individual fluxes in simulations with reduced SERCA revealed that the original step function caused LTCCs to play a larger role than expected. As SERCA function was restricted, flux through LTCCs increased on a scale that resulted in peak amplitude increasing with reduced SERCA solely because of the influx of \ca through LTCCs. This was deemed unlikely to be physiologically plausible and the voltage function reduced to its current form. The effects of this change $V$ on the base model are negligible.

\subsection*{Other fluxes across the SR membrane}
\begin{align}
	I_{\rm SERCA}&=\frac{g_{\rm SERCA}\cdot [{\rm Ca}^{2+}]_{i}^2}{K_{\rm SERCA}^2+[{\rm Ca}^{2+}]_{i}^2}\\
	I_{\rm SRl}&=g_{\rm SRl}\cdot\left([{\rm Ca}^{2+}]_{\rm SR}-[{\rm Ca}^{2+}]_{i}\right)
	\end{align}
	
\begin{align}
I_{\rm IP_3R}&=\frac{k_f\cdot N_{\rm IP_3R}}{V_{\rm myo}}\cdot P_{\rm IP_3R}\cdot\left([{\rm Ca}^{2+}]_{\rm SR}-[{\rm Ca}^{2+}]_{i}\right)\\
\end{align}
where, as described in \citet{sneyd_dynamical_2017}
\begin{align}
	P_{\rm IP_3R}=\frac{\beta}{\beta+k_\beta\cdot\left(\beta+\alpha\right)}
	\end{align}
Where $\alpha$ describes the rate of inactivation, $\beta$ the rate of activation. The parameter $k_\beta$ is used to fit to data, as in \citet{sneyd_dynamical_2017}
\begin{align}
	\alpha&=(1-B)\cdot\left(1-m\cdot h_\alpha\right)\\
\beta&=B\cdot m\cdot h
\end{align}
Here $B$ describes the dependence on \ipns, $m$ the dependence on \cans, and $h$ and its limit $h_\alpha$ is a delay factor which also has a dependence on \cans. In this study, we are primarily interested in the dependence of \ipr channels on \ca so we fix $B$ and focus on the other parameters.

\begin{align}
	m&=\frac{[{\rm Ca}^{2+}]_i^4}{K^4_c+[{\rm Ca}^{2+}]_i^4}\\
\frac{dh}{dt}&=\frac{\left(h_\alpha-h\right)\cdot\left(K^4_t+[{\rm Ca}^{2+}]_i^4\right)}{t_{max}\cdot K^4_t}\\
h_\alpha&=\frac{K^4_h}{K^4_h+[{\rm Ca}^{2+}]_i^4}
\end{align}
The values $m$ and $h_\alpha$ are Hill functions. Together they are controlled by two parameters which determine the \ca dependence of \ipr channels: $K_c$ and $K_h$. \ipr channels are active at the intersection of these two functions.
	
\subsection*{Cytosolic buffers}
\begin{align}
	I_{\rm TnC}&=k_{m_{\rm TnC}} \cdot \left(B_{\rm TnC}-{\rm TnC} \right) - k_{p_{\rm TnC}} \cdot {\rm TnC} \cdot [{\rm Ca}^{2+}]_{i}\\
	\beta_{\rm fluo} &= \left(1+\frac{K_{\rm fluo}\cdot B_{\rm fluo}}{\left( K_{\rm fluo}+[{\rm Ca}^{2+}]_{i}\right)^2}\right)^{-1}\\
	\beta_{\rm CaM} &= \left(1+\frac{K_{\rm CaM}\cdot B_{\rm CaM}}{\left(K_{\rm CaM}+[{\rm Ca}^{2+}]_{i}\right)^2}\right)^{-1}
\end{align}

\subsection*{\mynotes{NFAT Coupling}}
\mynotes{Total cellular NFAT is split between four states: nuclear phosphorylated, $A_{pn}$, and dephosphorylated, $A_n$, and cytosolic phosphorylated, $A_{pc}$, and dephosphorylated, $A_c$ and cycles between them.}
\mynotes{\begin{align}
    \frac{dA_n}{dt}&=J_2C_{cn}-J_3\\
    \frac{dA_{pn}}{dt}&=J_3-J_4\\
    \frac{dA_c}{dt}&=J_1-J_2\\
    \frac{dA_{pc}}{dt}&=\frac{J_4}{C_{cn}}-J_1
\end{align}}
\mynotes{$C_{cn}$ is a scaling factor that accounts for the difference in volume between the cytosol and the nucleus while $J_1$, $J_2$, $J_3$, and $J_4$ are flux terms that describe the movement of NFAT between these four states.}

\mynotes{\begin{align}
    J_1 &= k_{f,1}A_{pc}N_{\rm NFAT}f_{\rm CnA}-k_{r,1}A_c(1-f_{\rm CnA})\\
    J_2 &= k_{f,2}A_c\\
    J_3 &= k_{f,3}A_n(1-f_{\rm CnA})-k_{r,3}A_nN_{\rm NFAT}f_{\rm CnA}\\
    J_4 &= k_{f,4}A_{pn}
\end{align}}
Here $k_{f,i}$ are forward rate constants in the NFAT phosphorylation reaction, $k_{r,i}$ are reverse rate constants, and $f_{\rm CnA}$ is the fraction of activated calcineurin, given by
\mynotes{\begin{align}
    f_{\rm CnA}&=\frac{[{\rm Ca}^{2+}]^n_i}{[{\rm Ca}^{2+}]^n_i+K_{m,N}^n\left(1+\frac{K_{d,1}}{M}\right)}
\end{align}}
where $n$ is the Hill coefficient for calcineurin activation, $K_{m,N}$ is the half-maximal activation conentration of cytosolic \ca for calmodulin activation for the expected cellular concentration of calmodulin, and $K_{d,1}$ is the calcineurin-calmodulin dissociation constant. Parameter values for these constants can be found in Table \ref{tab:nfatparams}.

\section*{Code Availability}
CellML code is available on the Physiome Repository at:

https://models.physiomeproject.org/workspace/5ee \\

\noindent
Matlab code is available on github at:

https://github.com/CellSMB/compartmental\_ECC\_ETC

\section*{Parameter sensitivity analysis: Equations for main and total effects}
As described in \citet{saltelli_variance_2010}, the main and total effects were calculated by generating two sampling matrices $A$ and $B$ of model parameter values and then, for each parameter, a matrix $A^{(i)}_B$ for which all but the $i$th column match those of $A$ and the $i$th column is the $i$th column of $B$.

The main effect of parameter $i$ is then:
\begin{align}
    S_i &= V_{X_i}(E_{X_{\sim i}}(Y|X_i)) \Big/ V(Y)\\
    &= V(Y)-\sum^N_{j=1}\left(f(B)_j-f(A^{(i)}_B)_j\right)^2 \Big/ 2V(Y)N
\end{align}
The total effect of parameter $i$ is then:
\begin{align}
    S_{T_i} &= E_{X_{\sim i}}(V_{X_i}(Y|X_{\sim i})) \Big/ V(Y)\\
    &= \sum^N_{j=1}\left(f(A)_j-f(A^{(i)}_B)_j\right)^2 \Big/ 2V(Y)N
\end{align}
Here we denote $f(A)_j$ the results of the simulation with parameter in row $j$ of the sampling matrix $A$; $V(Y)$ the variance in simulation results across all rows of $A$ and $B$; and $N$ the number of rows in $A$ and $B$. 

In our parameter analysis, we generated parameter values within the range [0, 100] using the MATLAB sobolset function with Skip $1\times10^3$ and Leap $1\times10^2$, scrambled with the Mattousek-Affine-Owen algorithm. $N$ was set to $1\times10^6$.

\begin{table}%[tbhp]
\centering
\rowcolors{2}{White}{LightBlue!30}
\bf{Parameter values used in modelled $Ca^{2+}$ currents} \vspace{0.1cm}

\begin{tabular}[!ht]{@{}lll@{}}\toprule
	% \multicolumn{3}{r}{Global IP\textsubscript{3} ester} \\ \cmidrule(r){2-3}
	\textbf{Parameter} & 	\textbf{Description} & 	\textbf{Value}\\
	\midrule
	N & Number of CaRUs in the cell & 50 000 \\ \addlinespace
	V\textsubscript{myo} & Myocyte volume & 25.84$\times$10$^{3}$ $\mu$m$^3$ \\ \addlinespace
	N\textsubscript{IP\textsubscript{3}R} & Number of IP\textsubscript{3}R channels in the cell & 20 000 \\ \addlinespace
	g\textsubscript{SERCA} & Maximum pump rate of SERCA & 0.45 $\mu$M ms$^{-1}$ \\ \addlinespace
	K\textsubscript{SERCA} & Half saturation constant of SERCA & 0.5 $\mu$M \\ \addlinespace
	g\textsubscript{NCX} & Maximum pump rate of NCX & 38.5 $\mu$M ms$^{-1}$ \\ \addlinespace
	$\eta$ & Voltage dependence of NCX & 0.35 \\ \addlinespace
	K\textsubscript{mNa} & Na\textsuperscript{+} half saturation of NCX & 87.5 mM \\ \addlinespace
	K\textsubscript{mCa} & Ca\textsuperscript{2+} half saturation of NCX & 1.380 mM \\ \addlinespace
	k\textsubscript{sat} & Low potential saturation factor of NCX & 0.1 \\ \addlinespace
	g\textsubscript{PMCA} & Maximum pump rate of Ca$^{2+}$-ATPase & 3.5 nM ms\textsuperscript{-1}\\ \addlinespace
	K\textsubscript{PMCA} & Half saturation constant of Ca$^{2+}$-ATPase & 38.5 $\mu$M ms$^{-1}$ \\ \addlinespace
	g\textsubscript{CaB} & Conductance of background Ca\textsuperscript{2+} current & 2.32 $\times$10$^{-5}$ $\mu$M ms$^{-1}$ mV$^{-1}$ \\ \addlinespace
	g\textsubscript{SRl} & Pump rate of NCX & 1.8951 $\times$10$^{-5}$ ms$^{-1}$ \\ \addlinespace
	K\textsubscript{t} & \ipr delayed response parameter & 0.1 $\mu$M\\ \addlinespace
	t\textsubscript{max}&\ipr recovery time parameter & 1000 s$^{-1}$\\ \addlinespace
	\bottomrule
	% This differs between Hinch paper and cellML 
\label{tab:params1}
\end{tabular}
\caption{N\textsubscript{IP\textsubscript{3}R} from \citet{harzheim_increased_2009}. All other values from \cite{hinch_simplified_2004}}
\end{table}

\begin{table}%[tbhp]
\centering
\rowcolors{2}{White}{LightBlue!30}
\bf{Fixed ionic concentrations and buffer parameters}

\begin{tabular}[!ht]{@{}lcc@{}}\toprule
	% \multicolumn{3}{r}{Global IP\textsubscript{3} ester} \\ \cmidrule(r){2-3}
	\textbf{Ion} & 	\textbf{Description} & 	\textbf{Value}\\
	\midrule
	$[$Na\textsuperscript{+}$]$\textsubscript{i} & Intracellular sodium & 10 mM \\ \addlinespace
	$[$Na\textsuperscript{+}$]$\textsubscript{e} & Extracellular sodium  & 140 mM \\ \addlinespace
	$[$Ca\textsuperscript{2+}$]$\textsubscript{e} & Extracellular calcium  & 1 mM \\ \addlinespace
	B\textsubscript{CaM} & Total cytosolic concentration of calmodulin & 50 x 10\textsuperscript{-3} mM \\ \addlinespace
	K\textsubscript{CaM} & Half saturation constant of calmodulin & 2.38 x 10\textsuperscript{-3} mM \\ \addlinespace
	B\textsubscript{TnC} & Total cytosolic concentration of troponin & 70 x 10\textsuperscript{-3} mM\\ \addlinespace
	k\textsubscript{m\textsubscript{TnC}} & Dissociation rate of Ca\textsuperscript{2+} to troponin & 0.04\textsuperscript{3} mM$^{-1}$ ms$^{-1}$ \\ \addlinespace
	k\textsubscript{p\textsubscript{TnC}} & Binding rate of Ca\textsuperscript{2+} to troponin & 0.04 $\mu$M\textsuperscript{-1} ms$^{-1}$ \\ \addlinespace
	B\textsubscript{fluo} & Concentration of Fluo-4AM dye & 1 x 10\textsuperscript{-3} mM \\ \addlinespace
	K\textsubscript{fluo} & Dissociation constant of Fluo-4AM dye & 1 mM \\ \addlinespace
	V\textsubscript{0} & Resting membrane potential & $-$80 mV \\ \bottomrule
\label{tab:params2}
\end{tabular}
\caption{K\textsubscript{fluo} and B\textsubscript{fluo} from \cite{thomas_comparison_2000}. All other values from \cite{hinch_simplified_2004}}

% \addtabletext{nomenclature for the TSs refers to the numbered species in the table.}
\end{table}

\begin{table}%[tbhp]
\centering
\rowcolors{2}{White}{LightBlue!30}
\bf{Parameter values used to model NFAT4} \vspace{0.1cm}

\begin{tabular}[!ht]{@{}lll@{}}\toprule
	% \multicolumn{3}{r}{Global IP\textsubscript{3} ester} \\ \cmidrule(r){2-3}
	\textbf{Parameter} & 	\textbf{Description} & 	\textbf{Value}\\
	\midrule
	 $C_{cn}$ & Volume difference between nucleus and cytosol & 50 \\ \addlinespace
	$k_{f,1}$ & Rate constant & 7.69 $\times$ 10\textsuperscript{-6} nM\textsuperscript{-1}s\textsuperscript{-1} \\ \addlinespace
	$k_{f,2}$ & Rate constant & 1.44 $\times$ 10\textsuperscript{-3} s\textsuperscript{-1} \\ \addlinespace
	$k_{f,3}$ & Rate constant & 3.62 $\times$ 10\textsuperscript{-4} s\textsuperscript{-1} \\ \addlinespace
	$k_{f,4}$ & Rate constant & 4.45 $\times$ 10\textsuperscript{-4} s\textsuperscript{-1} \\ \addlinespace
	$k_{r,1}$ & Rate constant & 1.93 $\times$ 10\textsuperscript{-2} s\textsuperscript{-1} \\ \addlinespace
	$k_{r,3}$ & Rate constant & 4.71 $\times$ 10\textsuperscript{-5} nM\textsuperscript{-1}s\textsuperscript{-1} \\ \addlinespace
	$n$ & Hill coefficient of calcineurin & 2.92\\
	\addlinespace
	$K_{m,N}$ & Calcium-calcineurin constant & 535 nM\\
	\addlinespace
	$K_{d,1}$ & calcineurin-calmodulin dissociation constant & 1760 nM\\
	\addlinespace
	$M$ & Expected cellular calmodulin concentration & 6000 nM\\
	\addlinespace
	\bottomrule
	% This differs between Hinch paper and cellML 
\end{tabular}
\caption{\mynotes{Parameters for the parsimonious NFAT model by \citet{cooling_sensitivity_2009}}}
\label{tab:nfatparams}
\end{table}

\newpage
\section*{Supplementary Figures}

\begin{figure}[htp]
	\centering
	\raisebox{-0.8\height}{\includegraphics[scale=0.5]{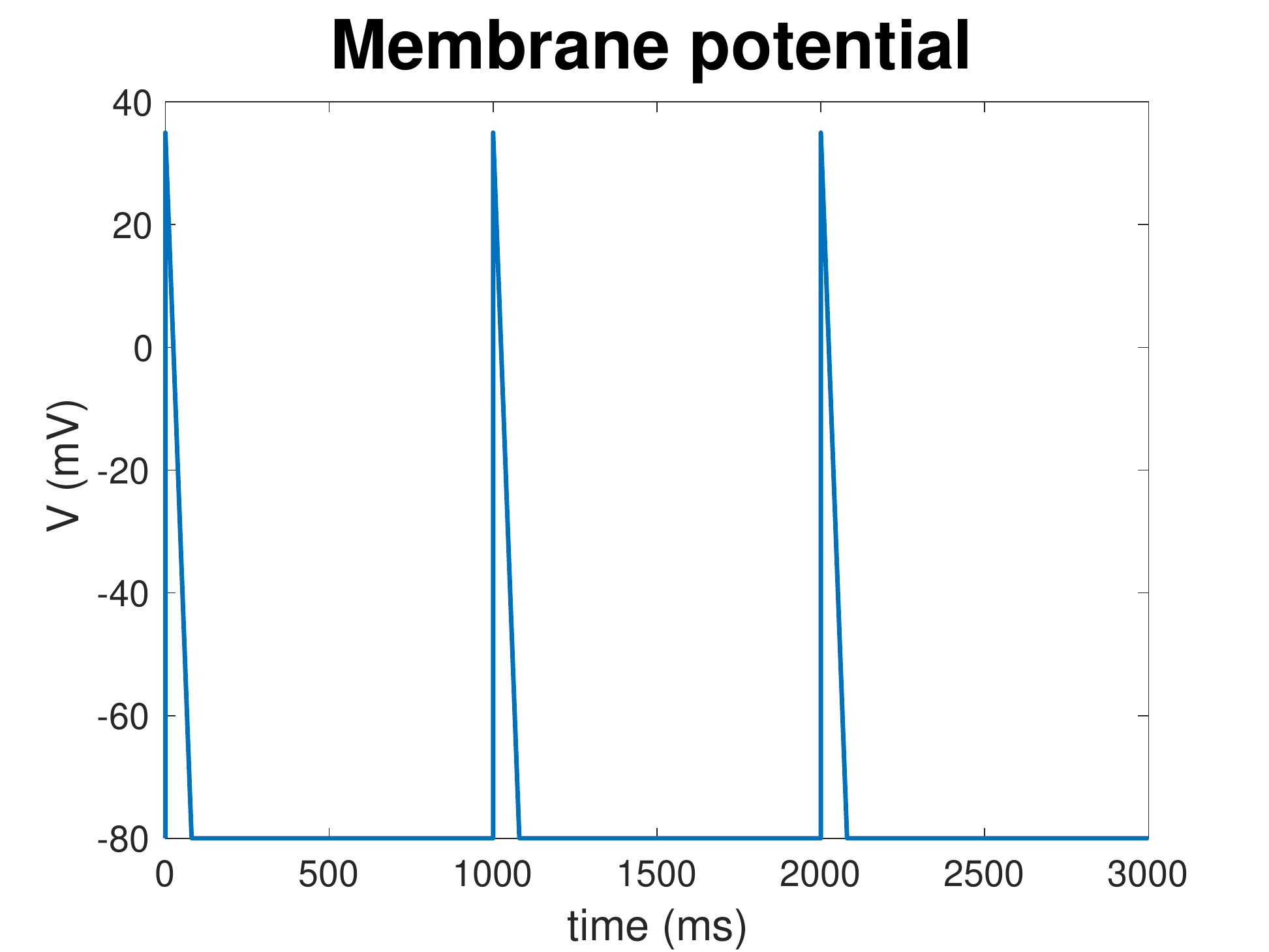}}
	\caption{Membrane depolarisation initiating each calcium transient.}
	\label{fig:dc}
\end{figure}

\begin{figure}[htp]
	\centering
	\begin{subfigure}[b]{0.45\textwidth}
		\includegraphics[width=1.\textwidth]{Figures/contour_ip3_K_c_Amplitude_0-3hzip3_new.eps}
		\put(-215,200){\Large A}
	\end{subfigure}
	\begin{subfigure}[b]{0.45\textwidth}\hspace{5mm}
		\includegraphics[width=1.\textwidth]{Figures/contour_ip3_K_c_FDHM_0-3hzip3_new.eps}
		\put(-215,200){\Large B}
	\end{subfigure}
	~\\
	\begin{subfigure}[b]{0.45\textwidth}
		\includegraphics[width=1.\textwidth]{Figures/contour_ip3_K_c_Baseline_0-3hzip3_new.eps}
		\put(-215,200){\Large C}
	\end{subfigure}
	\begin{subfigure}[b]{0.45\textwidth}\hspace{3mm}
		\includegraphics[width=0.86\textwidth]{Figures/cr_ip3_0-3Hz_t}
		\put(-180,200){\Large D}
	\end{subfigure}
    \caption{Effect of \ip concentration and the parameter $K_c$ on the calcium transient with large delay and pacing frequency of 0.3 Hz. These two parameters, along with maximum \ipr flux, have the greatest impact when considering the effect of \ipr activation on the calcium transient.}
	\label{fig:effect_kc_ip3_3Hz}
\end{figure}

\begin{figure}[ht]
	\centering
	\includegraphics[width=\textwidth]{Figures/fluxhipvaryingkf3}
	\caption{Simulated hypertrophic ECC transient and fluxes with varying $k_f$, $K_c$. The sign of $I_{NCX}$ indicates whether calcium is moving into (positive) or out of (negative) the cell.  Parameters here are chosen to show the system behaviour at each region illustrated in Figure 7. The crosses in each of Figures \ref{fig:effect_kc_ip3_3Hz} and 7 match the colours of the corresponding transients in this figure. \ip concentration is $10$ $\mu$M in all simulations. The model is paced at 0.3 Hz.}
	\label{fig:flux_vary_kf}
\end{figure}

\begin{figure}[ht]
	\centering
	\includegraphics[width=0.8\textwidth]{Figures/fluxhip_allvaryingIP3.eps}
	\caption{\mynotes{Simulated hypertrophic ECC transient and fluxes with varying [\ipns], $K_c$. The sign of $I_{NCX}$ indicates whether calcium is moving into (positive) or out of (negative) the cell.  Parameters here are chosen to show the system behaviour at each region illustrated in Figure 4. The crosses in each of Figures 4 and 6 match the colours of the corresponding transients in this figure. \ip concentration is $10$ $\mu$M in all simulations. The model is paced at 1 Hz.}}
	\label{fig:all_fluxes}
\end{figure}

\newpage
\clearpage
\bibliographystyle{biophysj}
\renewcommand\refname{Supporting References}
\bibliography{Hunt2020biblio.bib}